\documentclass[a4paper,11pt]{article}
\pdfoutput=1 

\usepackage{jheppub} 

\usepackage[T1]{fontenc} 

\usepackage{graphicx}
\usepackage{latexsym}
\usepackage{amsfonts}
\usepackage{amsmath}
\usepackage{amssymb}
\usepackage{revsymb}
\usepackage{bm}
\usepackage{extarrows}
\usepackage{subfigure}
\usepackage{simplewick}
\usepackage[vcentermath]{youngtab}
\usepackage{slashed}

\newcommand{\vev}[1]{\langle #1 \rangle}
\newcommand{\bvev}[1]{\bigl\langle #1 \bigr\rangle}

\newcommand{\VEV}[1]{\left\langle #1 \right\rangle}

\makeatletter
\newcommand{\xRightarrow}[2][]{\ext@arrow 0359\Rightarrowfill@{#1}{#2}}
\makeatother

\begin{document}

\title{$\bm{B}$-decay anomalies and scalar leptoquarks in unified Pati-Salam models from noncommutative geometry}

\author[a,1]{Ufuk Aydemir,\note{Corresponding author.}}
\author[b]{Djordje Minic,}
\author[c,d,e]{Chen Sun,}
\author[b]{and Tatsu Takeuchi}


\affiliation[a]{School of Physics, Huazhong University of Science and Technology,\\Wuhan, Hubei 430074, P. R. China}
\affiliation[b]{Center for Neutrino Physics, Department of Physics, Virginia Tech, Blacksburg, VA 24061 USA}
\affiliation[c]{CAS Key Laboratory of Theoretical Physics, Institute of Theoretical Physics,\\Chinese Academy of Sciences, Beijing 100190, P. R. China}
\affiliation[d]{Department of Physics, Brown University, Providence, RI, 02912, USA}
\affiliation[e]{Department of Physics and Astronomy, Dartmouth College, Hanover, NH 03755, USA}

\emailAdd{uaydemir@hust.edu.cn}
\emailAdd{dminic@vt.edu}
\emailAdd{chen.sun@brown.edu}
\emailAdd{takeuchi@vt.edu}

\abstract{
Motivated by possible scalar-leptoquark explanations of the recently reported $B$-decay anomalies, 
we investigate whether the required leptoquarks can be accommodated within
models based on noncommutative geometry (NCG).
The models considered have the gauge structure of Pati-Salam models,
$SU(4)\times SU(2)_L\times SU(2)_R$, 
with gauge coupling unification at a single scale.~In one of the models,
we find a unique scalar leptoquark with quantum numbers $(3,1,-\frac{1}{3})_{321}$, originating from a complex multiplet $(6,1,1)_{422}$, which can potentially explain the $B$-decay anomalies if its mass is on the
order of a few TeV. The unification of couplings can be realized with the inclusion of a single step of intermediate symmetry breaking. The scalar leptoquark under consideration does not contribute to proton decay due to the absence of diquark couplings, as dictated by the underlying noncommutative geometry.
}


\arxivnumber{1804.05844}

\maketitle
\section{Introduction}

In papers \cite{Aydemir:2015nfa} and \cite{Aydemir:2016xtj}
we have studied the phenomenology of models proposed by 
Chamseddine, Connes, and van Suijlekom in Refs.~\cite{Chamseddine:2013rta,Chamseddine:2015ata}, which are based on Connes' noncommutative geometry (NCG) \cite{Connes:1994yd,Connes:2007book}.
The models are characterized by a Pati-Salam (PS) gauge structure 
$G_{422}=SU(4)\times SU(2)_L \times SU(2)_R$
with the requirement that the gauge couplings unify at a single scale,
and a scalar sector whose content and couplings are fixed by the NCG of the models.
The fermion content is the same as the Standard Model (SM) plus the right-handed
neutrinos of each generation.

The unification of gauge couplings is a distinguishing requirement of NCG based versions of the SM \cite{Chamseddine:2010ud,Chamseddine:2012sw}, as well as its $G_{422}$ extensions discussed in this paper \cite{Chamseddine:2013rta,Chamseddine:2015ata}.
It is due to the underlying ``spectral action'' having only one overall coupling,
despite the fact that the gauge groups do not unify into a single simple Lie group,
that is, the model is not a Grand Unified Theory (GUT).
Note that coupling unification is not a requirement of canonical
non-GUT $G_{422}$ models found in the literature \cite{Mohapatra:2002book,Pati:1974yy,Mohapatra:1974gc,Mohapatra:1974hk,Senjanovic:1975rk}. 

Another feature of the NCG-PS framework is its restricted scalar
  content compared to the canonical Pati-Salam and GUT frameworks,
  which is appealing in terms of the predictability of the NCG
  framework. The scalar content of the three NCG-PS models proposed in
  Refs.~\cite{Chamseddine:2013rta,Chamseddine:2015ata} is listed in
  Table~\ref{NCG-HiggsContent} and their full SM decomposition can be
  inferred from Table~\ref{Decompositions}. As shown
    in Refs.~[3,4], the scalar
    sector is restricted due to the geometric feature of NCG
    as outlined in Section~4.1. In essence, as the Higgs
  field is the gauge field in the discrete direction of NCG space,
  requiring gauge invariance in the total NCG space puts a stronger restriction on the Higgs sector
  than requiring it only in Minkowski space, hence reducing the
  number of free parameters and possible 
  combination of terms in the Higgs potential.\footnote{The fermion masses in NCG framework, as in the SM, are input in the NCG framework, which are contained in the generalized Dirac operator that is used to construct the NCG Lagrangian. In both the spectral Standard Model and its Pati-Salam extension, the smallness of the neutrino masses can be addressed in the context of seesaw mechanism~\cite{Chamseddine:2010ud,Chamseddine:2012sw}.}

The RG running in the NCG-PS framework is treated adopting the usual effective field theory approach. The difference between the canonical and NCG based Pati-Salam models is in the restricted scalar content of the NCG-PS formalism. The underlying noncommutative geometry, which is interpreted as the classical background, is assumed to be relevant at the UV. The SM and PS structures are interpreted as emergent from the corresponding underlying noncommutative geometry.\footnote{It is currently unknown if there are non-local degrees of freedom due to the geometric nature of the framework that may change the RG running in a non-Wilsonian way.}

In Ref.~\cite{Aydemir:2015nfa}, we investigated whether a 2 TeV $W_R$ boson \cite{Aad:2014aqa,Aad:2015owa,Aad:2015ipg,Khachatryan:2014gha,Khachatryan:2015sja,Khachatryan:2016yji}
could be accommodated within the NCG models considered, 
and in Ref.~\cite{Aydemir:2016xtj} whether a 750 GeV
diphoton resonance \cite{CMS:2015dxe,ATLAS:2015diphoton} could be accommodated.
Due to constraints imposed on the models by the underlying NCG,
no freedom exists to adjust the models' particle content, except via decoupling those that 
exist from the renormalization group (RG) running of the gauge coupling constants by
rendering them heavy through symmetry breaking.
Consequently, though it is not difficult to find particles within the models
that can be identified with the putative $W_R$ or the diphoton resonance,
we have found that lowering their masses to TeV scales
while maintaining the coupling unification condition is highly non-trivial.
The hints of the $W_R$ and diphoton resonance at the LHC have since disappeared \cite{Sirunyan:2018pom,Aaboud:2017yyg,Meideck:2018wbw}, but our
analyses nevertheless underscore the rigidity, and consequently the predictability, of the
NCG based models.

In this paper, we continue our investigation and question whether the models could 
naturally accommodate a TeV-scale scalar leptoquark necessary to 
explain the $B$-decay anomalies observed at the LHC and elsewhere \cite{Ciezarek:2017yzh}.
The leptoquark of interest is $S_1(\overline{3},1,\frac{1}{3})_{321}$,
a colored isoscalar with electromagnetic charge $\frac{1}{3}$.
In order to explain the $B$-decay anomalies, its mass must be in the few TeV range
\cite{Bauer:2015knc,Becirevic:2016oho,Cai:2017wry,Freytsis:2015qca}.
Again, it is not difficult to identify scalars with the required quantum numbers
within the NCG models, but the quantum number assignment alone does not 
necessarily guarantee that the scalars couple to lepton-quark pairs as leptoquarks should do.
And again, the challenge is whether coupling unification can be achieved or not.
On this question, we recall that in the context of $SU(5)$ GUT, the presence of certain
leptoquarks actually helps in unifying the couplings without SUSY \cite{Murayama:1991ah,Cox:2016epl}.
Our study will focus on whether the field corresponding to the $S_1$ has a similar
effect in NCG models.

Another appealing aspect of the NCG-based Pati-Salam models, pointed out in this paper, is the proton stability. Ordinarily, in the usual Pati-Salam framework the stability of proton is not always guaranteed in the case of $S_1$ being light. In the NCG formalism, on the other hand, the proton stability is ensured thanks to the absence of $S_1$'s diquark couplings, as dictated by the underlying noncommutative geometry.

This paper is organized as follows: 
In section 2 we briefly review the current status of $B$-decay anomalies.
In section 3 we list the various leptoquark explanations that have been proposed in the literature, and identify the introduction of a single $S_1$ leptoquark as the most attractive
solution.  
Section 4 discusses the scalar content of the NCG models and their Yukawa couplings,
and we find a unique field that would serve our purpose as the $S_1$ leptoquark.
Section 5 presents an analysis on how and whether the coupling unification condition can be
maintained when the $S_1$ mass is a few TeV.
Section 6 concludes with a discussion of how the proton decay is, perhaps surprisingly, not a problem in our context, as well as some
concluding comments on the NCG framework and the current status of non-commutativity in quantum gravity and string theory. We end with an appendix in which we review the derivation of the
most generic Lagrangian for scalar and vector leptoquark
interactions with Standard Model (SM) fermions.



\section{$B$-decay Anomalies}

During the past few years, several disagreements between experiment and Standard Model (SM) predictions in rare $B$-decays have been observed by 
LHCb \cite{Aaij:2014ora,Aaij:2017vbb,RKstarnew,Aaij:2015yra,RDstarnew}, 
Belle \cite{Huschle:2015rga,Sato:2016svk,Hirose:2016wfn}, and
Babar \cite{Lees:2012xj,Lees:2013uzd}.
The decay channels in question are
$B^+\rightarrow K^+ \mu^+\mu^-$ \cite{Aaij:2014ora}, 
$B^0\rightarrow K^{*0} \mu^+\mu^-$ \cite{Aaij:2017vbb,RKstarnew},  and 
$\overline{B^0}\rightarrow D^{(*)+}\tau^-\overline{\nu}$ \cite{Lees:2012xj,Lees:2013uzd,Huschle:2015rga,Sato:2016svk,Hirose:2016wfn,Aaij:2015yra,RDstarnew}. 
Expressed in terms of the ratios
\begin{equation}
R_{K^{(*)}} \;=\; \dfrac{\mathcal{B}(B\rightarrow K^{(*)}\mu^+\mu^-)}{\mathcal{B}(B\rightarrow K^{(*)}e^+e^-)}
\end{equation}
LHCb reports \cite{Aaij:2014ora,Aaij:2017vbb,RKstarnew}\footnote{%
Previous measurements of these ratios by Belle and Babar 
were consistent with the SM within their larger experimental uncertainties.
}
\begin{eqnarray}
R_{K^{\phantom{*}}} & = & \;\;\,
0.745 \;{}^{+\,0.090}_{-\,0.074}\,(\mathrm{stat}) \pm 0.036\,(\mathrm{syst})\quad 
\mbox{for $1<q^2<6\;\mathrm{GeV^2/c^4}$}\;,
\vphantom{\Big|}
\\
R_{K^{*}} & = & 
\begin{cases}
0.660 \; {}^{+\,0.110}_{-\,0.070}\,(\mathrm{stat}) \pm 0.024\,(\mathrm{syst}) & \mbox{for $0.045 <q^2<1.1\;\mathrm{GeV^2/c^4}$}\\
0.685 \; {}^{+\,0.113}_{-\,0.069}\,(\mathrm{stat}) \pm 0.047\,(\mathrm{syst}) & \mbox{for $1.1<q^2<6.0\;\mathrm{GeV^2/c^4}$}\\
\end{cases}
\;,
\end{eqnarray}
where $q^2$ is the invariant mass of the lepton pair in the final state.
Here, the upper cut $q^2_{\max} = 6\;\mathrm{GeV^2/c^4}$ is imposed to avoid the effects
of the $J/\psi$ and higher $c\bar{c}$ resonances, 
while a common lower cut 
$q^2_{\min}$ for both electron and muon final states is imposed to avoid phase space effects.
In the case of $R_{K^*}$, the lower cut has been taken all the way down to
$q^2_{\min} = 0.045\;\mathrm{GeV^2/c^4} \approx 4m_\mu^2$, 
but the range from this $q^2_{\min}$ to $q^2_{\max} = 6\;\mathrm{GeV^2/c^4}$ has been divided into two
at $q^2 = 1.1\;\mathrm{GeV^2/c^4}$ in order to 
isolate the contribution of the $\phi$ resonance 
($B^0\to \phi(\to\ell^+\ell^-) K^{*0}$) into the lower bin.
With these bounds on $q^2$ in place, hadronic uncertainties 
have been argued to cancel in these ratios and that the SM predictions of $R_K$ and $R_{K^*}$ are unity with only $O(10^{-4})$ uncertainties \cite{Hiller:2003js,Bobeth:2007dw,Bouchard:2013mia,DAmico:2017mtc,Alok:2017sui}.
However, it has also been argued that both QED \cite{Bordone:2016gaq} and
QCD \cite{Capdevila:2017ert} corrections and uncertainties were underestimated in these predictions.
Ref.~\cite{Bordone:2016gaq} argues that a more realistic
set of SM predictions is
\begin{equation}
\begin{array}{ll}
R_K^{\mathrm{SM}}[1,6] & = \; 1.00\pm 0.01\;,\\
R_{K^{*}}^{\mathrm{SM}}[1.1,6] & = \; 1.00\pm 0.01\;,\\
R_{K^{*}}^{\mathrm{SM}}[0.045,1.1] & = \; 0.906\pm 0.028\;,
\end{array}
\end{equation}
while Ref.~\cite{Capdevila:2017ert} lists
\begin{equation}
\begin{array}{ll}
R_{K^{*}}^{\mathrm{SM}}[1.1,6] & = \; 1.000\pm 0.006\;,\\
R_{K^{*}}^{\mathrm{SM}}[0.045,1.1] & = \; 0.922\pm 0.022\;.
\end{array}
\end{equation}
Thus, the prediction that the values of $R_{K}^{\mathrm{SM}}[1,6]$ and $R_{K^{*}}^{\mathrm{SM}}[1.1,6]$ are unity appears to be fairly robust, and we find that
the experimental values of $R_K[1,6]$ and $R_{K^*}[1.1,6]$ are both 
suppressed compared to their SM predictions by 
$\sim 2.6\,\sigma$ \cite{Aaij:2014ora,Aaij:2017vbb,RKstarnew}.
%

For the semileptonic $B$ to $D$ decays, the ratios
\begin{eqnarray}
R_{D^{(*)}} & = & \dfrac{\mathcal{B}(B\rightarrow D^{(*)}\tau\nu)}{\mathcal{B}(B\rightarrow D^{(*)}\ell\nu)}\;,\qquad\qquad
\mbox{$\ell\;=\; e$ or $\mu$}\;,
\end{eqnarray}
have been measured using a variety of techniques by
Babar \cite{Lees:2012xj,Lees:2013uzd}, 
Belle \cite{Huschle:2015rga,Sato:2016svk,Hirose:2016wfn}, 
and
LHCb ($R_{D^*}$ only) \cite{Aaij:2015yra,RDstarnew}.
For the denominator of the above ratios,
Belle and Babar take the average of $B\to D^{(*)}e\nu$ and $B\to D^{(*)}\mu\nu$
branching fractions, while LHCb uses that for $B\to D^{(*)}\mu\nu$.
The world averages as of summer 2017, according to the Heavy Flavor Averaging Group \cite{HFAG:2017}, are
\begin{eqnarray}
R_{D^{\phantom{*}}} & = & 0.407 \pm 0.039\,(\mathrm{stat}) \pm 0.024\,(\mathrm{syst}) \;,\cr
R_{D^*} & = & 0.304 \pm 0.013\,(\mathrm{stat}) \pm 0.007\,(\mathrm{syst}) \;. 
\end{eqnarray}
The Standard Model (SM) predictions of these ratios, on the other hand, are \cite{Fajfer:2012vx,Aoki:2016frl}\footnote{See also the recent analyses in Ref.~\cite{Jaiswal:2017rve} and Ref.~\cite{Bigi:2017jbd}, both of which obtain slightly different values for these ratios.}
\begin{eqnarray}
R_D^{\mathrm{SM}} & = & 0.300\pm 0.008\;,\cr
R_{D^*}^{\mathrm{SM}} & = & 0.252 \pm 0.003\;,
\end{eqnarray}
and one sees that the experimental values 
$R_D$ and $R_{D^*}$ are in excess of their SM predictions by 
$2.3\,\sigma$ and $3.4\,\sigma$, respectively \cite{HFAG:2017}. 

These experimental results,\footnote{See also Ref.~\cite{Ciezarek:2017yzh} for a recent review.} taken at face value, challenge lepton flavor universality \cite{Glashow:2014iga}
and could be interpreted as signatures of new physics beyond the SM.
Note that for new physics to account for $R_{K^{(*)}} < R_{K^{(*)}}^{\mathrm{SM}}$ one could either suppress the numerator or enhance the denominator, while 
for $R_{D^{(*)}} > R_{D^{(*)}}^{\mathrm{SM}}$
one could either enhance the numerator or suppress the denominator.
However, any new physics involving the electron is generically highly 
constrained.
Therefore, new physics needs to suppress the process $B\to K^{(*)}\mu^+\mu^-$, 
while enhancing the process $B\to D^{(*)}\tau\nu_\tau$.

\section{Scalar Leptoquarks}


Unlike several other LHC ``anomalies'' that have come and gone \cite{Aaboud:2017yyg,Meideck:2018wbw,Sirunyan:2018pom}
the persistence of the $B$-decay anomalies has garnered a great deal of interest
and various explanations involving new physics have been proposed.
Here, we focus our attention to those involving scalar leptoquarks.\footnote{%
For a recent general review on leptoquark-related phenomenology, see Ref.~\cite{Dorsner:2016wpm}. The classic reference on leptoquarks is Ref.~\cite{Buchmuller:1986zs}. 
For completeness, we review the basic theoretical facts from these references in the Appendix.}

As reviewed in the Appendix, there exist six possible 
$G_{321}=SU(3)_C\times SU(2)_L\times U(1)_Y$ quantum number
assignments to scalar leptoquarks as listed in Table~\ref{LQtable}.
The corresponding six scalar leptoquark fields are labelled 
$S_1$, $\widetilde{S}_1$, $\overline{S}_1$, $S_3$, $R_2$, and $\widetilde{R}_2$.
Their couplings to the quarks and leptons are shown explicitly in 
Eqs.~(\ref{LQcomponentcouplings2}) and (\ref{LQcomponentcouplings0})
and also summarized in the last column of Table~\ref{LQtable}.

\begin{figure}[t]
\includegraphics[width=16cm]{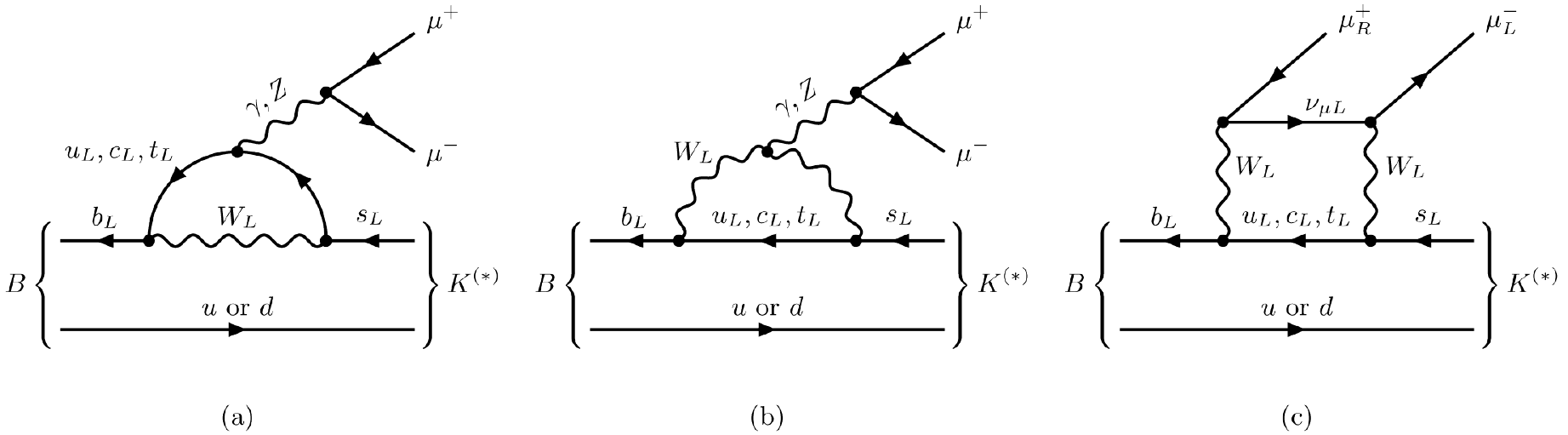}
\caption{\label{SMprocesses}
SM processes that contribute to $B\to K^{(*)}\mu^+\mu^-$.
The muons in the final states of (a) and (b) do not have definite chirality.
}
\end{figure}

\begin{figure}[t]
\includegraphics[width=16cm]{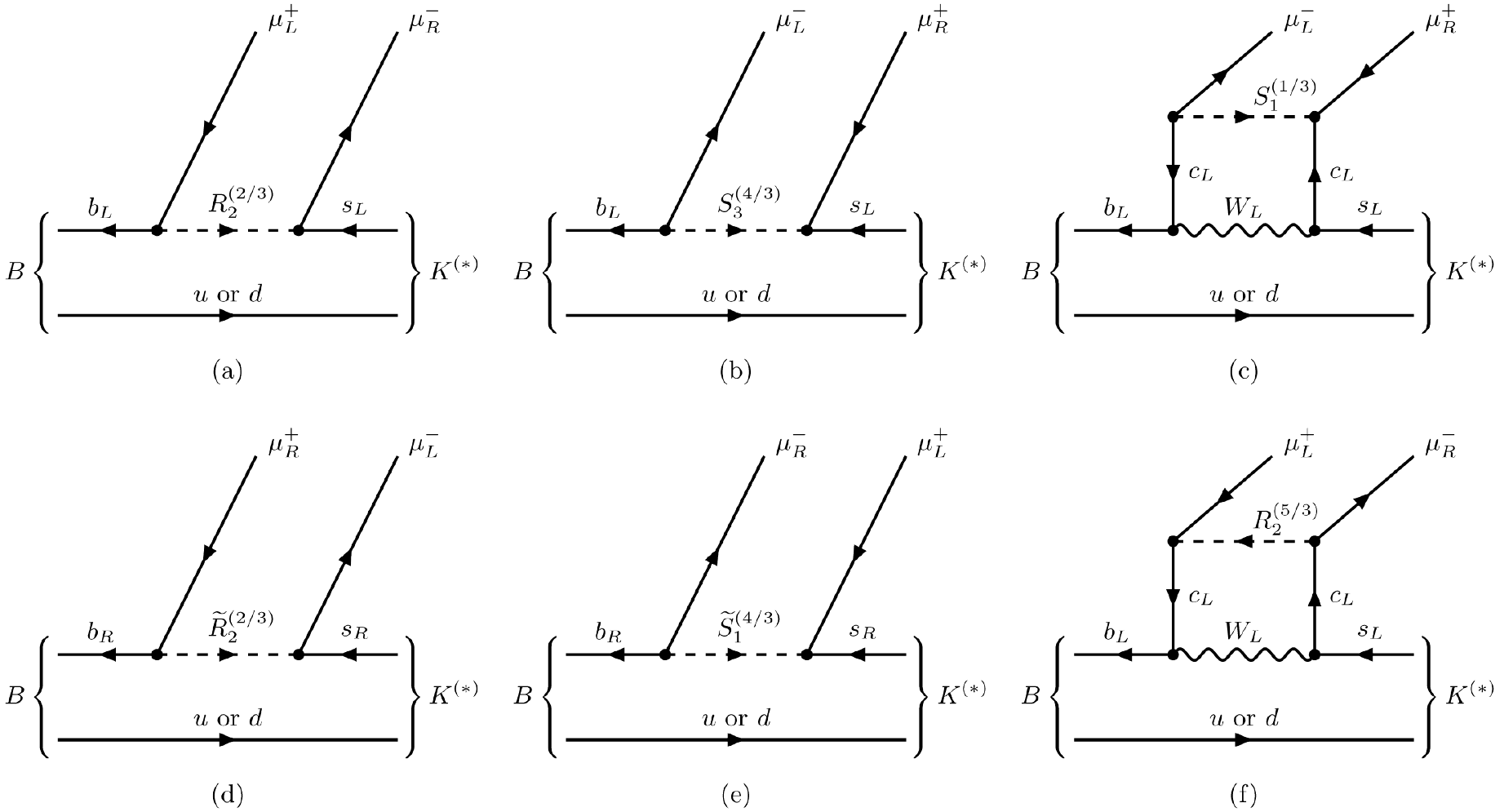}
\caption{\label{LQprocesses}
Possible leptoquark contributions to $B\to K^{(*)}\mu^+\mu^-$.}
\end{figure}

\begin{figure}[t]
\begin{center}
\includegraphics[width=12cm]{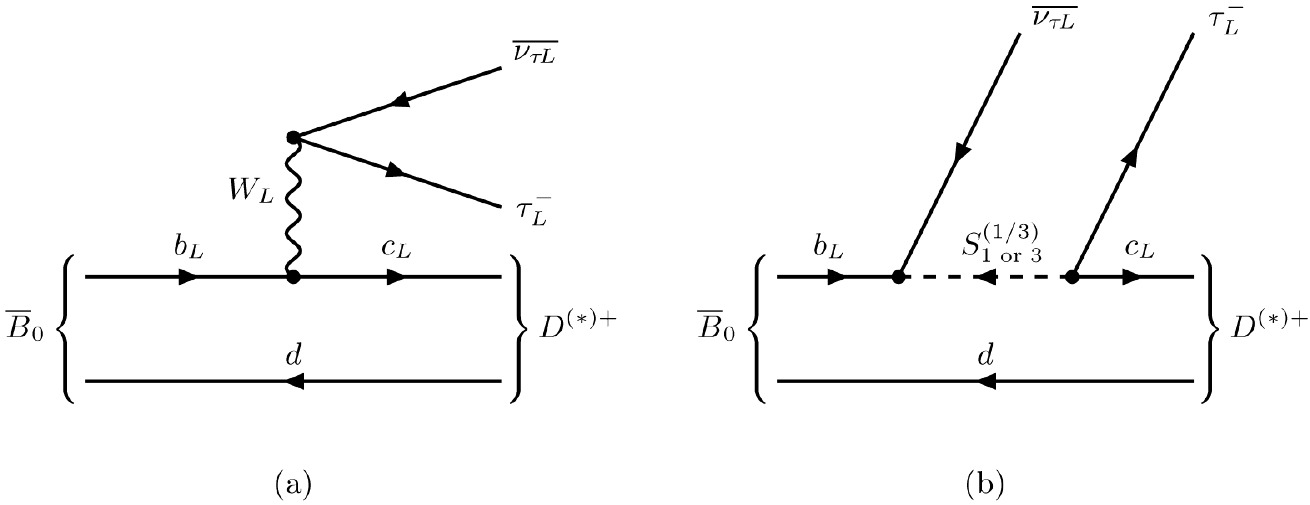}
\caption{\label{BtoKlnu}
(a) SM and (b) possible leptoquark contributions to 
$\overline{B}_0\to D^{(*)+}\tau^-\overline{\nu_{\tau\!L}}$.}
\end{center}
\end{figure}


To explain the $R_{K^{(*)}}$ anomalies one needs to induce 
$b_L\to s_L\mu^+\mu^-$ operators which can interfere destructively with
the SM processes shown in Figure~1.
These operators are denoted in the
literature as \cite{Grinstein:1988me,Misiak:1992bc,Buras:1994dj}\footnote{%
Ref.~\cite{Grinstein:1988me} uses a slightly different numbering scheme 
in which $\mathcal{O}_{9/10}$ are respectively denoted $\mathcal{O}_{8/9}$.
The normalizations of the operators also differ among various publications so care is
necessary when comparing the sizes of the Wilson coefficients.}
\begin{eqnarray}
\mathcal{O}_{9\phantom{0}}
& \propto &
\bigl(\overline{s_L}\gamma^\mu b_L\bigr)
\bigl(\overline{\mu}\gamma_\mu \mu\bigr)
\phantom{\gamma_5}
\;=\;
 2\bigl(\overline{s_L}\mu_R\bigr)\bigl(\overline{\mu_R}b_L\bigr)
+2\bigl(\overline{b_L^C}\mu_L\bigr)\bigl(\overline{\mu_L}s_L^C\bigr)
\;,
\vphantom{\Big|}
\cr
\mathcal{O}_{10}
& \propto &
\bigl(\overline{s_L}\gamma^\mu b_L\bigr)
\bigl(\overline{\mu}\gamma_\mu \gamma_5 \mu\bigr)
\;=\;
 2\bigl(\overline{s_L}\mu_R\bigr)\bigl(\overline{\mu_R}b_L\bigr)
-2\bigl(\overline{b_L^C}\mu_L\bigr)\bigl(\overline{\mu_L}s_L^C\bigr)
\;.
\vphantom{\Big|}
\end{eqnarray}
Looking through the leptoquark-quark-lepton couplings listed in
Eqs.~(\ref{LQcomponentcouplings2}) and (\ref{LQcomponentcouplings0}),
we can see that in order to induce these operators at tree-level one needs 
$S_3^{(4/3)}$, which induces $\mathcal{O}_{9} - \mathcal{O}_{10}$ 
\cite{Hiller:2014yaa,Aloni:2017ixa,DAmico:2017mtc}, or
$R_2^{(2/3)}$, which induces $\mathcal{O}_{9} + \mathcal{O}_{10}$ 
\cite{Becirevic:2016oho,Becirevic:2017jtw,DAmico:2017mtc}.
See Figures~2(a) and 2(b).
The wrong chirality operators
\begin{eqnarray}
\mathcal{O}_{9\phantom{0}}^{\prime}
& \propto &
\bigl(\overline{s_R}\gamma^\mu b_R\bigr)
\bigl(\overline{\mu}\gamma_\mu \mu\bigr)
\phantom{\gamma_5}
\;=\;
 2\bigl(\overline{b_R^C}\mu_R\bigr)\bigl(\overline{\mu_R}s_R^C\bigr)
+2\bigl(\overline{s_R}\mu_L\bigr)\bigl(\overline{\mu_L}b_R\bigr)
\;,
\vphantom{\Big|}
\cr
\mathcal{O}_{10}^{\prime}
& \propto &
\bigl(\overline{s_R}\gamma^\mu b_R\bigr)
\bigl(\overline{\mu}\gamma_\mu\gamma_5 \mu\bigr)
\;=\;
 2\bigl(\overline{b_R^C}\mu_R\bigr)\bigl(\overline{\mu_R}s_R^C\bigr)
-2\bigl(\overline{s_R}\mu_L\bigr)\bigl(\overline{\mu_L}b_R\bigr)
\;,
\vphantom{\Big|}
\end{eqnarray}
which contribute to $b_R\to s_R\mu^+\mu^-$, are obtained from the exchange of 
$\widetilde{R}_2^{(2/3)}$, which induces $\mathcal{O}_{9}^{\prime} - \mathcal{O}_{10}^{\prime}$ \cite{Hiller:2014yaa,Cox:2016epl,Becirevic:2015asa,Becirevic:2016oho,Becirevic:2016yqi,DAmico:2017mtc,Queiroz:2014pra}, 
or
$\widetilde{S}_1^{(4/3)}$, which in turn induces $\mathcal{O}_{9}^{\prime} + \mathcal{O}_{10}^{\prime}$ \cite{Hiller:2014yaa,DAmico:2017mtc}.
See Figures~2(d) and 2(e).
Global fits that have been performed by various groups, e.g. Refs.~\cite{Descotes-Genon:2015uva,Mahmoudi:2016mgr,Altmannshofer:2017fio,DAmico:2017mtc,Alok:2017sui},
indicate that
a suppression of the Wilson coefficient $C_9$ (the coefficient of $\mathcal{O}_9$)
compared to its SM value by about 25\% provides the best fit.

For the $R_{D^{(*)}}$ anomalies we need an operator which 
interferes constructively with the SM process
shown in Figure~3(a).
The required operator is
\begin{eqnarray}
\bigl(\overline{c_L}\gamma^\mu b_L\bigr)
\bigl(\overline{\tau_L} \gamma_\mu \nu_{\tau L}\bigr)
& = & 
\bigl(\overline{b_L^C}\gamma^\mu c_L^C\bigr)
\bigl(\overline{\tau_L} \gamma_\mu \nu_{\tau L}\bigr)
\;=\;
2\bigl(\overline{b_L^C}\nu_{\tau L}^{\phantom{C}}\bigr)
 \bigl(\overline{\tau_L}c_L^C\bigr)
\;.
\end{eqnarray}
To induce this at tree-level one needs 
$S_1^{(1/3)}$ \cite{Bauer:2015knc,Freytsis:2015qca,Becirevic:2016oho,Cai:2017wry,Crivellin:2017zlb,Marzocca:2018wcf,Buttazzo:2017ixm} or 
$S_3^{(1/3)}$ \cite{Crivellin:2017zlb,Marzocca:2018wcf,Buttazzo:2017ixm} which couple to \textit{left-handed} quarks and leptons. 
See Figure~3(b).
We find the introduction of the $S_1^{(1/3)}$ leptoquark 
coupled to left-handed fermions particularly attractive
since, as pointed out in Ref.~\cite{Bauer:2015knc}, 
in addition to explaining the $R_{D^{*}}$ anomalies at tree level, 
it can also explain the $R_{K^{(*)}}$ anomalies at the one-loop level, the same level as the SM contributions.\footnote{See Refs.~\cite{Crivellin:2017zlb,Marzocca:2018wcf,Buttazzo:2017ixm} for possible two-leptoquark solutions that utilize $S_1(\overline{3},1,\frac{1}{3})_{321}$ and $S_3(\overline{3},3,\frac{1}{3})_{321}$.}  See Figure~2(c).
Under closer scrutiny, it was suggested in Ref.~\cite{Becirevic:2016oho} that the 
$S_1^{(1/3)}$ model may not be able to explain $R_{K^{(*)}}$ without coming into conflict 
with the experimental values of the ratios 
\begin{equation}
R_{D^{(*)}}^{\mu/e} \;=\; 
\dfrac{\mathcal{B}(B\to D^{(*)}\mu\nu)}
      {\mathcal{B}(B\to D^{(*)}e\nu)}\;.
\end{equation}
However, Ref.~\cite{Cai:2017wry} argues that this constraint can be
circumvented by allowing the $S_1^{(1/3)}$ mass to be a TeV or larger,
the preferred mass being a few TeV.

Assuming that an $S_1^{(1/3)}$ leptoquark coupled to left-handed fermions
with a mass of a few TeV can simultaneously explain the $R_{K^{(*)}}$ and $R_{D^{(*)}}$ anomalies, 
we inquire whether such a leptoquark can be accommodated within the
NCG models of Chamseddine et al. \cite{Chamseddine:2013rta,Chamseddine:2015ata}.

\section{Scalar Leptoquarks in unified Pati-Salam models from NCG}

In this section, we list the three NCG-based unified Pati-Salam-like models proposed by
Chamseddine, Connes, and van Suijlekom in Refs.~\cite{Chamseddine:2013rta,Chamseddine:2015ata}, and specify how leptoquarks can be fit into their particle contents.
We begin with a brief outline of how NCG models are constructed, 
in particular, the three models of Refs.~\cite{Chamseddine:2013rta,Chamseddine:2015ata}.

\subsection{NCG Model Construction}
\label{sec:ncg-model-constr}


Noncommutative geometry (NCG) generalizes the concepts of spinor bundle and
principle bundle, and manages to combine the two concepts into one in a
mathematically rigorous fashion~\cite{Connes:1994yd}. 
As a result, the Dirac operator has both a continuous part and a discrete part.
The continuous part realizes the ordinary Dirac operator, and the
discrete part accounts for the representation space of the gauge algebra. 

Consequently, in this framework, the most general gauge transformation includes both
a transformation along the continuous directions, 
which is identified with the usual gauge transformation,
and that in the discrete direction, which is not explicit in the
usual field theoretical formalism. 
To maintain gauge invariance of the whole action, 
one needs gauge fields in both directions, which
are interpreted as the ordinary gauge fields (continuous direction) and the Higgs
fields (discrete direction). 

It is worth noting that if one tries to interpret the space as an
extra-dimension model, the metric dimension of the extra dimension is zero, because the extra direction is discrete.
Therefore, in contrast to the usual extra-dimension models,
no extra effort is required to stabilize the two ``sheets.'' 
%
%
%
%
In what follows, we show this more concretely.
The continuous gauge fields $A_\mu$ transform in the usual way: 
%
\begin{eqnarray}
  \label{eq:continuous-gauge}
A_\mu'
& = &
u\, A_\mu^{\phantom{\prime}} u^{-1} - \frac{i}{g}\, u\, \partial_\mu u^{-1}
\cr
& = &
A_\mu + u\left[A_\mu, u^{-1}\right] - \frac{i}{g} u \left[ \partial_\mu, u^{-1}\right] 
\;, 
\end{eqnarray}
where $u$ is an element of the gauge group,
while the discrete part, the Higgs field $\Phi$, transforms in an analogous fashion with $\partial_\mu$
replaced by a matrix commutator $[D, \; \cdot \; ]$, which represents
a ``derivative'' in the discrete direction\footnote{
Apparently,
  Leibniz rule is satisfied by the commutator. One might think that
  $[D, \; \cdot \; ]$ cannot work as a derivation as it is
  not nilpotent in general. However, in the context of cycles over a
  $C^*$ algebra where integrals are defined using Hochschild cochains,
  $D$ does work as a derivation.
  For details on how to
  calculate such integrals,
  c.f. Ref.~\cite{Connes:1994yd,Connes:2007book}. 
}:
\begin{equation}
\label{eq:discrete-derivative}
\Phi'
\;=\;
\Phi + u\left[\Phi, u^{-1}\right] - \frac{i}{g} u\left[D, u^{-1}\right]\;.
\end{equation}
%
Here, the matrix $u$ is still a transformation 
generated by the gauge algebra, 
and the matrix $D$ is the discrete part of the Dirac operator. In NCG,
the gauge field is defined as as the summation $ \sum a[\mathcal{D}, b]$,
with $a,b$ being elements of the gauge
algebra, and $\mathcal{D}$ is the generalized Dirac
operator~\cite{Connes:2007book}. Putting $\mathcal{D}$ equal $\slashed
\partial$ (or $D$), we can reproduce all properties of $\slashed A$ (or $\Phi$), respectively. Therefore, the last terms of 
Eqs.~\eqref{eq:continuous-gauge} and \eqref{eq:discrete-derivative} can be
combined with the gauge field that is being transformed, which
guarantees that the result of the transformation on a gauge field
is still a gauge field. On the other hand, depending on the symmetry group,
one may or may not be able to write the $u[\Phi,u^{-1}]$ term
in Eq.~\eqref{eq:discrete-derivative}
into the $a[D, b]$ form. 
In other words, due to
the presence of the $u[\Phi,u^{-1}]$ term, the form of a gauge field defined as $\sum a[D,b]$ may not be
closed after a gauge transformation.   To maintain the gauge invariance, high
order terms of the gauge field, arising from $u[\Phi, u^{-1}]$, are
needed to build a covariant derivative. The requirement of vanishing $u[\Phi, u^{-1}]$
is the so called order-one condition \cite{Connes:2007book}.


It has been argued \cite{Connes:2007book,Chamseddine:2013rta} that the
requirement of the order-one condition, together with a  Majorana
mass term for right handed neutrinos, selects the SM gauge group from
$G_{422} = SU(4) \times SU(2)_L \times SU(2)_R$. 
Furthermore, it has been shown \cite{Chamseddine:2013rta} that if one starts with the 
$G_{422} = SU(4) \times SU(2)_L \times SU(2)_R$ gauge group, 
and (partially) lifts the order-one condition, up to three models can be constructed. 
To be specific, starting with $G_{422} = SU(4) \times SU(2)_L \times SU(2)_R$, 
if the order-one condition is imposed on $G_{321} = SU(3)_C \times SU(2)_L \times U(1)_Y$, 
\textit{i.e.} if the SM gauge transformation does not generate
any higher order terms from $U[\Phi, U^{-1}]$ that cannot be written in the
$\sum a[D, b]$ form, one ends with Model A of Ref.~\cite{Chamseddine:2015ata}; 
if the order-one condition is lifted and if one
only looks into a left-right asymmetric subset of the moduli space of the Dirac operator, 
one has Model B of Ref.~\cite{Chamseddine:2015ata}; 
if the order-one condition is lifted and the full moduli space of the Dirac
operator is considered, the result is Model C of Ref.~\cite{Chamseddine:2015ata}.


We note in passing that, if one starts with $G_{422} = SU(4) \times SU(2)_L \times SU(2)_R$, 
it has been observed in \cite{Chamseddine:2015ata} that the breaking of this gauge symmetry
can be achieved either from usual Higgs mechanism as in Refs.~\cite{Aydemir:2015nfa,Aydemir:2016xtj}, or from the use of the
order-one condition \cite{Chamseddine:2007hz,Chamseddine:2013rta}. 
While it is possible that the the order-one condition encodes the symmetry breaking pattern, 
we refrain from discussing how the two seemingly different approaches could be related at some fundamental level to achieve similar results.

\subsection{The three models}

\begin{table}[t]
\begin{center}
\begin{tabular}{l|l|ll}
\hline
{\small Model} & {\small Symmetry} & {\small Higgs Content} \vphantom{\Big|}\\
\hline\hline
A & $G_{422}$  & $\phi(1,2,2)_{422}$, $\Sigma(15,1,1)_{422}$, & \hspace{-3mm}$\widetilde{\Delta}_R(4,1,2)_{422}$  \vphantom{\bigg|}\\
\hline
B & $G_{422}$  & $\phi(1,2,2)_{422}$, $\widetilde{\Sigma}(15,2,2)_{422}$, & \hspace{-3mm}$\Delta_R(10,1,3)_{422}$, $H_R(6,1,1)_{422}$ \vphantom{\bigg|}\\
\hline
C & $G_{422D}$ & $\phi(1,2,2)_{422}$, $\widetilde{\Sigma}(15,2,2)_{422}$, & \hspace{-3mm}$\Delta_R(10,1,3)_{422}$, $H_R(6,1,1)_{422}$, \vphantom{\bigg|}\\
  & & & \hspace{-3mm}$\Delta_L(10,3,1)_{422}$, $H_L(6,1,1)_{422}$ \vphantom{\Big|}\\
\hline\hline
$SO(10)$ & $G_{422}$ &  $\phi(1,2,2)_{422}$, $\Sigma(15,1,1)_{422}$, & \hspace{-3mm}$\Delta_R(10,1,3)_{422}$ \vphantom{\bigg|}\\
\hline
\end{tabular}
\end{center}
\caption{\label{NCG-HiggsContent}
The scalar content of the three NCG based unified $G_{422}$ models
proposed by Chamseddine, Connes, and van Suijlekom in Refs.~\cite{Chamseddine:2013rta} and \cite{Chamseddine:2015ata}.
The last row lists for comparison the scalar content of the $SO(10)$ based $G_{422}$ model 
discussed in Ref.~\cite{Aydemir:2016qqj}, below its unification scale where 
the $SO(10)$ symmetry is broken to $G_{422}$. Note that in our notation throughout the paper the subscripts $L$ and $R$ of scalars indicate the chirality of fermions that the corresponding scalars couple to, which is shown explicitly
in Section~4 through the decomposition of the interaction terms, such as Eqs.~(\ref{DeltaRcouplings1}), (\ref{HRcouplings}), (\ref{DeltaLcouplings1}), and (\ref{HLcouplings}).}
\end{table}

The three NCG models proposed in Refs.~\cite{Chamseddine:2013rta,Chamseddine:2015ata},
which we refer to as models A, B, and C,
differ in their scalar sector content, and their unbroken symmetry structure is listed in Table~\ref{NCG-HiggsContent}\footnote{Recently, it has been argued that the incorporation of the Clifford structures in the spectral action formalism of the noncommutative geometry yields additional scalars as well~\cite{Kurkov:2017wmx}. The resulting model in that case turns out to be the Standard Model augmented by several scalars fields that carry the quantum numbers of the leptoquarks $\overline{S}_1$ and $\widetilde{R}_2$.}.
As in our previous work \cite{Aydemir:2015nfa, Aydemir:2016xtj} we use the following notation for the symmetries:
\begin{eqnarray}
G_{422D} & = & SU(4)_C\otimes SU(2)_L\otimes SU(2)_R\otimes D\;,\vphantom{\Big|}\cr
G_{422}  & = & SU(4)_C\otimes SU(2)_L\otimes SU(2)_R\;,\vphantom{\Big|}\cr
G_{3221} & = & SU(3)_C\otimes SU(2)_L\otimes SU(2)_R\otimes U(1)_{B-L}\;,\vphantom{\Big|}\cr
G_{321}  & = & SU(3)_C\otimes SU(2)_L\otimes U(1)_{Y}\;,\vphantom{\Big|}\cr
G_{31}   & = & SU(3)_C\otimes U(1)_{em} \;,\vphantom{\Big|}
\label{Gdef}
\end{eqnarray}
where $D$ in $G_{422D}$ refers to the left-right symmetry, a $Z_2$ symmetry which keeps the left and the right sectors equivalent. 
We adopt the normalization of the hypercharge $Y$ so that 
$Q_{em} = I_3^L + Y = I_3^L + I_3^R + (B-L)/2$.
For comparison, the last row of Table~\ref{NCG-HiggsContent} lists the
scalar content of an $SO(10)$ based $G_{422}$ model studied in
Ref.~\cite{Aydemir:2016qqj}, after $SO(10)$ is broken to $G_{422}$.



When the $G_{422}$ (models A and B and $SO(10)$) or $G_{442D}$ (model C) of the NCG 
based models break to $G_{321}$ of the SM, the scalars listed in Table~\ref{NCG-HiggsContent}
decompose into irreducible representations of $G_{321}$ as listed in the third column of Table~\ref{Decompositions}.
Of the fields listed there, all the color-triplet fields have the quantum numbers of 
leptoquarks.  For instance, the $\Delta_{R}(4,1,2)_{422}$ field which appears in
model~A decomposes as
\begin{eqnarray}
\Delta_{R}(4,1,2)_{422}
& \xrightarrow{G_{422}\to G_{3321}} &
 \widetilde{\Delta}_{R1}(1,1,2,-1)_{3221} 
+\widetilde{\Delta}_{R3}\left(3,1,2,\frac{1}{3}\right)_{3221}
\cr 
& \xrightarrow{G_{3221}\to G_{321}} & 
\biggl[ \widetilde{\Delta}_{R1}(1,1,0)_{321} 
+\widetilde{\Delta}_{R1}(1,1,-1)_{321}
\biggr]
\cr
& & 
+\biggl[\,
 \underbrace{\widetilde{\Delta}_{R3}\left(3,1,\frac{2}{3}\right)_{321}}_{\displaystyle \overline{S}_1^*}
+\underbrace{\widetilde{\Delta}_{R3}\left(3,1,-\frac{1}{3}\right)_{321}}_{\displaystyle S_1^*}
\,\biggr]
\;,
\end{eqnarray}
and we see that the two colored fields in the last line have 
quantum numbers corresponding to leptoquarks $\overline{S}_1$ and $S_1$.
Fields with quantum numbers corresponding to the six leptoquarks 
$S_1$, $\widetilde{S}_1$, $\overline{S}_1$,
$S_3$, $R_2$, and $\widetilde{R}_2$ all occur in one model or another.
However, for any of these fields to be identifiable as leptoquarks, 
they must couple to quark-lepton pairs after symmetry breaking.
In the following, we look at the scalar sectors of the three models one by one
and identify the fields that can be considered leptoquarks.
In particular, we will search for a field that can be identified as $S_1$ coupled to
left-handed fermions.

\subsection{Couplings between Fermions and Scalars}


Following Refs.~\cite{Chamseddine:2013rta,Chamseddine:2015ata}, 
we denote the $SU(2)_L$ and $SU(2)_R$ indices in the fundamental representation 
respectively with un-dotted and dotted lower-case Latin letters toward
the beginning of the alphabet: e.g.
$a = 1,2$ and $\dot{a} = 1,2$.
Note that despite their appearance, these are NOT spinor indices, and
so complex conjugation does not take on or off dots from the indices.\footnote{%
See, for instance, Ref.~\cite{Takeuchi:1900zz}.}
The $SU(4)$ index in the fundamental representation
is denoted with upper-case Latin letters toward the middle of the alphabet: e.g. $I=0,1,2,3$,
where $I=0$ is the lepton index and $I=i=1,2,3$ are the quark-color indices.\footnote{%
The authors of Refs.~\cite{Chamseddine:2013rta,Chamseddine:2015ata}
use $I=1,2,3,4$ to label the $SU(4)$ index, and then use $I=(1,i)$, with $i=1,2,3$ to
distinguish between the leptons and quarks. We instead use $I=0,1,2,3$, which seems more self-evident to us.
}
The fermion content of the models is
\begin{eqnarray}
\psi_{aI} & = & (\psi_{a0},\psi_{ai}) \;=\; 
\left( \begin{array}{ll} \psi_{10}, & \psi_{1i} \\ \psi_{20}, & \psi_{2i} \end{array} \right) \;=\;
\left( L_L, Q_L \right) \;=\;
\left( \begin{array}{ll} \nu_L, & u_L \\ e_L, & d_L \end{array} \right)
\;,\cr
\psi_{\dot{a}I} & = & (\psi_{\dot{a}0},\psi_{\dot{a}i}) \;=\; 
\left( \begin{array}{ll} \psi_{\dot{1}0}, & \psi_{\dot{1}i} \\ \psi_{\dot{2}0}, & \psi_{\dot{2}i} \end{array} \right) \;=\;
\left( L_R, Q_R \right) \;=\;
\left( \begin{array}{ll} \nu_R, & u_R \\ e_R, & d_R \end{array} \right)
\;,
\end{eqnarray}
that is, the SM content plus the right-handed neutrinos of each generation.
The generation and spinor indices are suppressed.
Complex (hermitian, Dirac) conjugation raises or lowers both indices, e.g.
\begin{equation}
\overline{\psi}{}^{aI} \;=\; \overline{\psi_{aI}} \;,\qquad
\overline{\psi}{}^{\dot{a}I} \;=\; \overline{\psi_{\dot{a}I}} \;.
\end{equation}
In the case of the $SU(2)$'s, the index can be lowered or raised using
\begin{equation}
(\epsilon)_{ab}\;,\qquad
(\epsilon^\dagger)^{ab}\;,\qquad
(\epsilon)_{\dot{a}\dot{b}}\;,\qquad
(\epsilon^\dagger)^{\dot{a}\dot{b}}\;,
\end{equation}
where $\epsilon = i\sigma_2$.

The most general $G_{422}$ invariant Yukawa interaction in the NCG models involving $\psi_{aI}$ and $\psi_{\dot{a}I}$ can be written schematically as
\begin{equation}
\label{eq:ncg-yukawa}
\mathcal{L}_\mathrm{Y}
\;=\;
\Bigl(\,
\overline{\psi}{}^{\dot{a}I}
\gamma_5 \Sigma^{bJ}_{\dot{a}I\vphantom{\dot{b}}} 
\psi_{bJ\vphantom{\dot{b}}}^{\vphantom{C}}
+ \overline{\psi^C}{}_{aI\vphantom{\dot{b}}}\gamma_5 H^{aIbJ} \psi_{bJ\vphantom{\dot{b}}}^{\vphantom{C}}
+ \overline{\psi^C}{}_{\dot{a}I\vphantom{\dot{b}}}\gamma_5 H^{\dot{a}I \dot{b}J} \psi_{\dot{b}J}^{\vphantom{C}}
\,\Bigr)
\;+\; h.c.\;,
\end{equation}
where $\psi^C = C\overline{\psi}^T$,
and the couplings constants are embedded in the complex scalar fields
$\Sigma^{bJ}_{\dot{a}I}$,
$H^{aIbJ}$, and $H^{\dot{a}I \dot{b}J}$.
The $\gamma_5$ that appears in this expression 
is due to {
the geometry being of even parity. The Hilbert space is $Z_2$
graded with $\gamma_5$ being the grading operator, so that the Dirac operator is
of odd degree. This is crucial in order to interpret the discrete
gauge field as the Higgs field. It is worth noting that, although seemingly originating
from a different assumption, the superconnection formalism also captures
this feature, as is shown in \cite{Aydemir:2014ama}. }


%
Since
\begin{equation}
\overline{\psi^C_{1}}\gamma_5\psi_{2}^{\phantom{C}} 
\;=\; \overline{\psi^C_{2}}\gamma_5\psi_{1}^{\phantom{C}}
\end{equation}
for any pair of anti-commuting fermionic operators $\psi_1$ and $\psi_2$,
the $H^{aIbJ}$ and $H^{\dot{a}I \dot{b}J}$ 
fields are respectively symmetric under the interchange of the indices
$(aI)\leftrightarrow(bJ)$ and $(\dot{a}I)\leftrightarrow(\dot{b}J)$.

As is discussed above, 
the way scalars couple to fermions is dictated by the generalized gauge invariance. 
{
In Eq.~\eqref{eq:ncg-yukawa},  
$\Sigma_{\dot{a}I}^{bJ}$ is the `connection' that links chiral fermions
to the ones with opposite chirality, \textit{i.e.}
it couples an $SU(2)_L$ fermion to an $SU(2)_R$ fermion. If a
fermion is mapped to the same fermion with opposite chirality, a Dirac
mass term can be generated after symmetry breaking. 
On the other hand,  $H_{\dot{a}I\dot{b}J}$ and $H_{aIbJ}$
 link fermions to anti-fermions with the same chirality. When a
 chiral fermion is mapped to its own charged conjugate, a Majorana
 mass term can be generated after symmetry breaking. 
}
Therefore, the former gives us $LR$ type coupling while the latter produces $RR$ and $LL$ types of coupling. This is how the Yukawa interaction terms are generated as a result of generalized gauge invariance. In particular, the NCG dictates there are only three terms, as indicated in Eq.~\eqref{eq:ncg-yukawa}.

The complex scalar fields $\Sigma^{bJ}_{\dot{a}I}$, $H_{\dot{a}I \dot{b}J}$, 
and $H_{aIbJ}^{\phantom{\dagger}}$ can, 
in general, consist of the following $G_{422}$ representations:
\begin{eqnarray}
\Sigma^{bJ}_{\dot{a}I}
& = & (1,2,2)_{422} + (15,2,2)_{422} \;, \vphantom{\Big|}\cr
H_{aIbJ}^{\phantom{\dagger}} & = & (6,1,1)_{422} + (10,3,1)_{422} \;, \vphantom{\Big|}\cr
H_{\dot{a}I \dot{b}J} & = & (6,1,1)_{422} + (10,1,3)_{422} \vphantom{\Big|}\;. 
\end{eqnarray}
Note that $6$ of $SU(4)$ and $1$ of $SU(2)_L$ ($SU(2)_R$) are respectively
anti-symmetric under $I\leftrightarrow J$ and $a\leftrightarrow b$ ($\dot{a}\leftrightarrow\dot{b}$)
rendering the $(6,1,1)_{422}$ representation symmetric under $(aI)\leftrightarrow(bJ)$ ($(\dot{a}I)\leftrightarrow(\dot{b}J)$).
Note also that the leptoquark $S_1$ we seek must couple to left-handed quarks and leptons,
so we can expect to find such a field embedded in the complex scalar field $H_{aIbJ}$.

Although the $G_{422}$ representations listed above are most general, 
and all the fields are contained in the model C, the models A and B
restrict the scalar sector to the fields listed in Table~\ref{NCG-HiggsContent}.
Since Model C is the only model which includes the field $H_{aIbJ}$, it is clear that
Model C is the place where we should be looking for our $S_1$ leptoquark. 
However, for the sake of completeness, we
look at the couplings of the scalar fields that appear in all three models,
 and identify all the fields that correspond to one type of leptoquark or another.

\subsubsection{Model A}
\label{sec:modelA}

In this model, the scalar field $H_{aIbJ}^{\phantom{\dagger}}$ is suppressed while the fields 
$\Sigma^{bJ}_{\dot{a}I}$ and $H_{\dot{a}I \dot{b}J}$ are decomposed as
\begin{eqnarray}
\label{eq:combination-couple-to-fermions}
\Sigma_{\dot{a}I}^{bJ}\,
& = &
  \Bigl( k^\nu \phi_{\dot{a}}^b 
       + k^e \,\widetilde{\phi}{}_{\dot{a}}^b 
  \Bigr) \Sigma_{I}^J / \Lambda
+ \Bigl( k^u \phi^{b}_{\dot{a}}
       + k^d \,\widetilde{\phi}_{\dot{a}}^b 
  \Bigr)
  \Bigl( \delta_I^J - \Sigma_I ^ J / \Lambda
  \Bigr)\;, 
\cr
H_{\dot{a}I\dot{b}J}
& = & \;\; k^{*\nu_R} \Delta_{\dot{a}J\vphantom{\dot{b}}} \Delta_{\dot{b}I} / \Lambda\;,
\vphantom{\Big|} 
\end{eqnarray}
where
%
\begin{equation}
\widetilde{\phi}{}_{\dot{a}}^{b} \;=\; (\epsilon)_{\dot{a}\dot{c}}\,\overline{\phi}{}^{\dot{c}}_{d}\,(\epsilon^\dagger)^{db}\;,\qquad
\epsilon\;=\; i\sigma_2\;.
\end{equation}
Note the manifest symmetry of $H_{\dot{a}I\dot{b}J}$ under $(\dot{a}I)\leftrightarrow(\dot{b}J)$.
The fields $\phi_{\dot a}^b$, $\Sigma_J^I$, $\Delta_{\dot{a}J}$
are those labeled as $\phi(1,2,2)_{422}$, $\Sigma(15,1,1)_{422}$, and $\widetilde{\Delta}_R(4,1,2)_{422}$
in Tables~\ref{NCG-HiggsContent} and \ref{Decompositions}. Also,
$k^\nu$, $k^e$, $k^u$, and $k^d$ denote Yukawa coupling matrices for the
neutrinos, charged leptons, up-type quarks, and down-type quarks, respectively, while
$k^{\nu_R}$ is the Majorana coupling matrix which gives Majorana masses to the right-handed neutrinos upon symmetry breaking, and the asterisk stands for complex conjugation.
In order to give mass-dimension one to all scalar fields, we have introduced the scale $\Lambda$
which is only implicit in Refs.~\cite{Chamseddine:2013rta,Chamseddine:2015ata}.
%

Note that the above decompositions render the interactions of the scalar fields 
$\widetilde{\Delta}_R(4,1,2)_{422}$ and $\Sigma(15,1,1)_{422}$ and the fermions into mass-dimension five operators, \textit{i.e.}
the generalized version of the $LHLH$ operator of the SM.
To generate dimension-four operators of the Yukawa or Majorana type, we must give vacuum 
expectation values (VEV's) to the scalar fields.
In the current case, VEV's are given to  
\begin{equation}
\label{eq:vev-connes}
\bvev{\Sigma_J^I}
\,=\,
\Lambda\,\delta^I_0\,\delta^0_J\;, 
\qquad
\bvev{\Delta_{\dot a J}}
\,=\,
w\,\delta^{\dot{1}}_{\dot{a}}\,\delta^{0}_{J}\;,
\qquad
\bvev{\phi_{\dot a}^b}
\,=\,
v\,\delta^{\dot{1}}_{\dot{a}}\,\delta^{b}_{1}\;.
\end{equation}
or in the notation of Table~\ref{Decompositions},
\begin{equation}
\bvev{\Sigma_1(1,1,0)_{321}} \,=\, \Lambda\,,
\quad
\bvev{\widetilde{\Delta}_{R1}(1,1,0)_{321}} \,=\, w\,,
\quad
\VEV{ \phi'_2\left(1,2,-\frac{1}{2}\right)_{321} } =
\left[\begin{array}{c} v \\ 0 \end{array}\right]\,.
\label{ModelA-VEVs}
\end{equation}
Note that $\widetilde{\phi}_{\dot{a}}^{b}$ and $\phi_{\dot{a}}^{b}$ in 
Eq.~(\ref{eq:combination-couple-to-fermions}) are not independent so 
$\vev{\phi_{\dot a}^b} = v\,\delta^{\dot{1}}_{\dot{a}}\,\delta^{b}_{1}$
implies 
$\vev{\widetilde{\phi}_{\dot a}^b} = v\,\delta^{\dot{2}}_{\dot{a}}\,\delta^{b}_{2}$.
The VEV's $\Lambda$ and $w$ break $SU(4)\times SU(2)_R$ to $SU(3)\times U(1)_Y$, 
while $v$ breaks $SU(2)_L\times U(1)_Y$ down to $U(1)_{em}$. 
So symmetry breaking generates the following mass terms for the fermions:
\begin{eqnarray}
\lefteqn{
\left[
  v\Bigl( \overline{\psi}{}^{\dot{1}0} k^\nu \psi_{10\vphantom{\dot{b}}}^{\vphantom{C}} 
       + \overline{\psi}{}^{\dot{2}0} k^e   \psi_{20\vphantom{\dot{b}}}^{\vphantom{C}}
  \Bigr) 
+ 
  v\Bigl( \overline{\psi}{}^{\dot{1}i} k^u   \psi_{1i\vphantom{\dot{b}}}^{\vphantom{C}}
       + \overline{\psi}{}^{\dot{2}i} k^d   \psi_{2i\vphantom{\dot{b}}}^{\vphantom{C}}
  \Bigr) 
+ \dfrac{w^2}{\Lambda}\,\overline{\psi^C}{}_{\dot{1}0}k^{\nu_R}\psi_{\dot{1}0}
\right]
\;+\; h.c.
} 
\cr
& = & 
\bigg[
  v\Bigl( \overline{\nu_R}\, k^\nu \nu_L
       + \overline{e_R}\, k^e e_L
  \Bigr) 
+ 
  v\Bigl( \overline{u_R}\, k^u u_L
       + \overline{d_R}\, k^d d_L
  \Bigr)
+ \dfrac{w^2}{\Lambda}\,\overline{\nu_R^C}\,k^{\nu_R} \nu_R^{\vphantom{\dagger}}
\bigg]
\;+\; h.c.
\cr
& &
\end{eqnarray}

The couplings of the $\phi(1,2,2)_{422}$, $\widetilde{\Delta}_R(4,1,2)_{422}$, and $\Sigma(15,1,1)_{422}$
fields to the fermions can be obtained by substituting 
Eq.~(\ref{eq:combination-couple-to-fermions}) into the generic expression Eq.~(\ref{eq:ncg-yukawa}),
taking into account Eq.~(\ref{eq:vev-connes}).
%
%
%
%
%
First, let us take 
$\Delta_{\dot{a}I} = \widetilde\Delta_R(4,1,2)_{422}$ as an example:
\begin{eqnarray}
\lefteqn{\overline{\psi^C}_{\dot{a}I\vphantom{\dot{b}}}\gamma_5 H^{\dot{a}I\dot{b}J} \psi_{\dot{b}J}
\;+\; h.c. \vphantom{\Big|}} 
\cr 
& = &
\dfrac{1}{\Lambda}\;
\overline{\psi^C}_{\dot{a}I\vphantom{\dot{b}}} 
\Delta^{\dot{b}I}
k^{\nu_R} 
\Delta^{\dot{a}J\vphantom{\dot{b}}}
\psi_{\dot{b}J} 
\;+\; h.c.
\vphantom{\bigg|}
\cr
& = & \dfrac{1}{\Lambda} 
\bigg[
\Bigl(
\overline{L_R^C} \widetilde{\Delta}_{R1}^{*\vphantom{C}} 
\Bigr)
k^{\nu_R}
\Bigl(
\widetilde{\Delta}_{R1}^{*\vphantom{C}} L_R^{\vphantom{C}} 
\Bigr)
+
\Bigl(
\acontraction{}{\overline{Q_R^C}}{\widetilde{\Delta}_{R1}^{*\vphantom{C}}\Bigr)k^{\nu_R}\Bigl(}{\widetilde{\Delta}_{R3}^{*\vphantom{C}}}
\overline{Q_R^C} \widetilde{\Delta}_{R1}^{*\vphantom{C}} 
\Bigr)
k^{\nu_R}
\Bigl(
\widetilde{\Delta}_{R3}^{*\vphantom{C}} L_R^{\vphantom{C}}
\Bigr) 
\cr
& & \quad
\;+\;
\Bigl(
\acontraction{\overline{L_R^C}}{\widetilde{\Delta}_{R3}^{*\vphantom{C}}}{\Bigr)k^{\nu_R}\Bigl(\widetilde{\Delta}_{R1}^{*\vphantom{C}}}{Q_R^{\vphantom{C}}}
\overline{L_R^C} \widetilde{\Delta}_{R3}^{*\vphantom{C}} 
\Bigr)
k^{\nu_R}
\Bigl(
\widetilde{\Delta}_{R1}^{*\vphantom{C}} Q_R^{\vphantom{C}}
\Bigr) 
+
\Bigl(
\acontraction[3pt]{}{\overline{Q_R^C}}{\widetilde{\Delta}_{R3}^{*\vphantom{C}}\Bigr)k^{\nu_R}\Bigl(}{\widetilde{\Delta}_{R3}^{*\vphantom{C}}}
\acontraction[7pt]{\overline{Q_R^C}}{\widetilde{\Delta}_{R3}^{*\vphantom{C}}}{\Bigr)k^{\nu_R}\Bigl(\widetilde{\Delta}_{R3}^{*\vphantom{C}}}{Q_R^{\vphantom{C}}}
\overline{Q_R^C} \widetilde{\Delta}_{R3}^{*\vphantom{C}} 
\Bigr)
k^{\nu_R}
\Bigl(
\widetilde{\Delta}_{R3}^{*\vphantom{C}} Q_R^{\vphantom{C}} 
\Bigr) 
\bigg]
\;+\; h.c.\;,
\end{eqnarray}
where the lines connecting the colored fields indicate color contraction, 
and we have also used the shorthand
\begin{equation}
\begin{array}{lll}
\widetilde{\Delta}_{R1} &\;=\; 
\widetilde{\Delta}_{R1}\left(1,1,2,-1\right)_{3221} &\;=\;
\left[
\begin{array}{l}
\widetilde{\Delta}_{R1}(1,1,\phantom{-}0)_{321} \\
\widetilde{\Delta}_{R1}(1,1,-1)_{321}
\end{array}
\right]
\;,\\
\widetilde{\Delta}_{R3} &\;=\;
\widetilde{\Delta}_{R3}\left(3,1,2,\frac{1}{3}\right)_{3221} &\;=\;
\left[
\begin{array}{l}
\widetilde{\Delta}_{R3}\left(3,1,+\frac{2}{3}\right)_{321} \\
\widetilde{\Delta}_{R3}\left(3,1,-\frac{1}{3}\right)_{321}
\end{array}
\right]
\;,
\end{array}
\end{equation}
for the fields that $\widetilde{\Delta}(4,1,2)_{422}$ decomposes into as
$G_{422}\to G_{3221}\to G_{321}$.
After the SM singlet $\widetilde{\Delta}_{R1}(1,1,0)_{321}$ develops a VEV,
Eq.~(\ref{ModelA-VEVs}), we find
\begin{eqnarray}
\lefteqn{\overline{\psi^C}_{\dot{a}I\vphantom{\dot{b}}}\gamma_5 H^{\dot{a}I\dot{b}J} \psi_{\dot{b}J}
\;+\; h.c. \vphantom{\bigg|}} 
\cr 
& \to & \dfrac{2w}{\Lambda}
\biggl[\;
\dfrac{w}{2}\!\left(\overline{\nu_R^C}\,k^{\nu_R}\,\nu_R^{\vphantom{\dagger}}\right)
+ \left(\overline{\nu_R^C}\,k^{\nu_R}\,\nu_R^{\vphantom{\dagger}}\right)\!\widetilde{\Delta}_{R1}^{'*(0)}
+ \left(\overline{\nu_R^C}\,k^{\nu_R}\,e_R^{\vphantom{\dagger}}  \right)\!\widetilde{\Delta}_{R1}^{*(1)}
\cr
& & \qquad\qquad\quad
+ \left(\overline{u_R^C}\,k^{\nu_R}\,\nu_R^{\vphantom{\dagger}}\right)\!
  \underbrace{\widetilde{\Delta}_{R3}^{*(-2/3)}}_{\displaystyle \overline{S}_1^{(-2/3)}}
+ \left(\overline{u_R^C}\,k^{\nu_R}\,e_R^{\vphantom{\dagger}}\right)\!
  \underbrace{\widetilde{\Delta}_{R3}^{*(1/3)}}_{\displaystyle S_1^{(1/3)}}
+ \cdots
\biggr]
\;+\; h.c.
\end{eqnarray}
where the first term inside the brackets is the Majorana mass term for the right-handed neutrinos,
the ellipses represent dimension 5 operators, and we have used the symbols 
\begin{equation}
\begin{array}{lll}
\widetilde{\Delta}_{R1}^{'*(0)} & = & \left[\widetilde{\Delta}_{R1}(1,1,0)_{321}\right]^* - w\;,
\vphantom{\Big|}\\
\widetilde{\Delta}_{R1}^{*(1)} & = & \left[\widetilde{\Delta}_{R1}(1,1,-1)_{321}\right]^* \;,\vphantom{\bigg|}\\
\widetilde{\Delta}_{R3}^{*(-2/3)} & = & \left[\widetilde{\Delta}_{R3}\left(3,1,+\frac{2}{3}\right)_{321}\right]^* \;,\vphantom{\Big|}\\
\widetilde{\Delta}_{R3}^{*(1/3)} & = & \left[\widetilde{\Delta}_{R3}\left(3,1,-\frac{1}{3}\right)_{321}\right]^* \;.\vphantom{\bigg|}
\end{array}
\end{equation}
Thus, we find that $\widetilde{\Delta}_R(4,1,2)_{422}$ includes the leptoquarks $S_1$ and $\overline{S}_1$,
but this $S_1$, by construction, only couples to $u_R$ and $e_R$.
Note also that it does not couple to diquarks.

Next, let us look at the couplings of $\phi_{\dot{a}}^{b} = \phi(1,2,2)_{422}$ and $\Sigma_I^J = \Sigma(15,1,1)_{422}$:
\begin{eqnarray}
\lefteqn{\overline{\psi}{}^{\dot{a}I} \gamma_5 \Sigma^{bJ}_{\dot{a}I}\psi_{bJ}^{\vphantom{I}}
\;+\; h.c.
\vphantom{\bigg|}}\cr
& = &
\overline{\psi}{}^{\dot{a}I} \gamma_5
\biggl[
  \Bigl( k^\nu \phi_{\dot{a}}^b 
       + k^e \,\widetilde{\phi}{}_{\dot{a}}^b 
  \Bigr) \Sigma_{I}^{J} / \Lambda
+ \Bigl( k^u \phi^{b}_{\dot{a}}
       + k^d \,\widetilde{\phi}_{\dot{a}}^b 
  \Bigr)
  \Bigl( \delta_{I}^{J} - \Sigma_{I}^{J} / \Lambda
  \Bigr)
\biggr] 
\psi_{bJ}
\;+\; h.c.
\;.
\end{eqnarray}
%
%
%
Note that if $k^u=k^\nu$ and $k^d=k^e$, then $\Sigma_I^J$ will decouple from the
fermions.
%
%
We are interested in the couplings of $\Sigma_I^J = \Sigma(15,1,1)_{422}$ to the fermions
so we replace $\phi_{\dot{a}}^{b}$ and $\widetilde{\phi}_{\dot{a}}^{b}$ with their VEV's,
and denote the shifted $\Sigma(15,1,1)_{422}$ fields as
\begin{equation}
\begin{array}{lll}
\Sigma_0^0 & = \; \Sigma_{1}\left(1,1,0\right)_{321} & =\; \Sigma_1^{(0)} \;=\; \Lambda + \Sigma_1^{'(0)} \;,\\
\Sigma_i^0 & = \; \Sigma_{3}\left(3,1,+\frac{2}{3}\right)_{321} & =\; \Sigma_{3}^{(2/3)}\;, \\
\Sigma_0^j & = \; \Sigma_{\overline{3}}\left(\overline{3},1,-\frac{2}{3}\right)_{321} & =\; \Sigma_{\overline{3}}^{(-2/3)}\;, \\
\Sigma_i^j & = \; \Sigma_{8}\left(8,1,0\right)_{321} & =\; \Sigma_{8}^{(0)}\;. 
\end{array}
\end{equation}
We find:
\begin{eqnarray}
\lefteqn{\overline{\psi}{}^{\dot{a}I} \gamma_5
\biggl[
  \Bigl( k^\nu \phi_{\dot{a}}^b 
       + k^e \,\widetilde{\phi}{}_{\dot{a}}^b 
  \Bigr) \Sigma_{I}^{J} / \Lambda
+ \Bigl( k^u \phi^{b}_{\dot{a}}
       + k^d \,\widetilde{\phi}_{\dot{a}}^b 
  \Bigr)
  \Bigl( \delta_{I}^{J} - \Sigma_{I}^{J} / \Lambda
  \Bigr)
\biggr] 
\psi_{bJ}
\;+\; h.c.
}
\cr
& \to &
-v 
\biggl[
  \Bigl( \overline{\psi}{}^{\dot{1}I} k^\nu \psi_{1J}
       + \overline{\psi}{}^{\dot{2}I} k^e   \psi_{2J}
  \Bigr) \Sigma_{I}^J / \Lambda
+ \Bigl( \overline{\psi}{}^{\dot{1}I} k^u   \psi_{1J}
       + \overline{\psi}{}^{\dot{2}I} k^d   \psi_{2J}
  \Bigr)
  \Bigl( \delta_I^J - \Sigma_{I}^{J} / \Lambda
  \Bigr)
\biggr] 
\;+\; h.c.
\phantom{\Bigg|}
\cr
& & = -v\biggl[
\Bigl( \overline{\nu_R}\,k^\nu \nu_L
       + \overline{e_R}\,k^e e_L
\Bigr)
+
\Bigl( \overline{u_R}\,k^u   u_L
       + \overline{d_R}\,k^d   d_L
\Bigr)
\biggr]
\vphantom{\Bigg|}
\cr 
& & \quad + \dfrac{v}{\Lambda}
\biggl[ 
  \Bigl\{ \overline{\nu_R}(\delta k^+) \nu_L
       + \overline{e_R}(\delta k^-) e_L
  \Bigr\} \Sigma^{'(0)}_{1}
+ \Bigl\{ \overline{u_R}(\delta k^+) u_L
       + \overline{d_R}(\delta k^-) d_L
  \Bigr\} \Sigma_{8}^{(0)}
\vphantom{\Bigg|}
\cr 
& & \quad\qquad
+ \Bigl\{ \overline{\nu_R}(\delta k^+) u_L
        + \overline{e_R}(\delta k^-) d_L
  \Bigr\}
  \Sigma_{\overline{3}}^{(-2/3)}
+ \Bigl\{ \overline{u_R}(\delta k^+) \nu_L
       + \overline{d_R}(\delta k^-) e_L
  \Bigr\}
  \Sigma_{3}^{(2/3)}
\biggr]
\cr
& & \quad
+\; h.c.\;,\vphantom{\Big|}
\end{eqnarray}
where
\begin{equation}
\delta k^+ \;\equiv\; k^u - k^\nu\;,\qquad
\delta k^- \;\equiv\; k^d - k^e\;.
\end{equation}
Comparing with Eq.~(\ref{LQcomponentcouplings0}), we can see 
that though $\Sigma_I^J = \Sigma(15,1,1)_{422}$ does not include fields corresponding to the
leptoquark doublets $R_2$ and $\widetilde{R}_2$, its components 
$\Sigma_3^{(2/3)}$ and $\Sigma_{\overline{3}}^{(-2/3)}$
do correspond to the following leptoquark components after the symmetry breaking $G_{321}\to G_{31}$:
\begin{equation}
\begin{array}{llll}
\Sigma_{3}^{(2/3)}             & \;\;\leftrightarrow\;\; & R_2^{(2/3)}\;,  & \widetilde{R}_2^{(2/3)} \;,\\
\Sigma_{\overline{3}}^{(-2/3)} & \;\;\leftrightarrow\;\; & R_2^{*(2/3)}\;, & \widetilde{R}_2^{*(2/3)} \;,
\end{array}
\end{equation}
with their couplings determined by the Yukawa coupling matrices $\delta k^+ = k^u - k^\nu$ and 
$\delta k^- = k^d - k^e$\;.
These fields also do not couple to diquarks.

\subsubsection{Model B}
\label{sec:model-b}

In Model B, the field $H_{aIbJ}^{\phantom{\dagger}}$ is still suppressed,
whereas the fields $\Sigma_{\dot{a}I}^{bJ}$ and $H_{\dot{a}I\dot{b}J}$ 
are no longer decomposed as shown in
Eq.~\eqref{eq:combination-couple-to-fermions}.
They are treated as mass-dimension one operators,
\textit{i.e.} `fundamental' in the language of Ref.~\cite{Chamseddine:2013rta}.

Although $H_{\dot{a}I\dot{b}J}$ is treated as a dimension-one field, 
it is still a product representation 
$(4,1,2) \times (4,1,2) = (10, 1, 3) + (6,1,1)$.
Recall the symmetry of $H_{\dot{a}I\dot{b}J}$ under $(\dot{a}I)\leftrightarrow(\dot{b}J)$.
We can write
\begin{eqnarray}
H_{\dot{a}I\dot{b}J} 
& = & \underbrace{\dfrac{1}{4}\left(
        H_{\dot{a}I\dot{b}J}
      + H_{\dot{b}I\dot{a}J}
      + H_{\dot{b}J\dot{a}I} 
      + H_{\dot{a}J\dot{b}I}
      \right)}_{\displaystyle = \Delta_{(\dot{a}\dot{b})(IJ)} = \Delta_R(10,1,3)_{422}}
    + \underbrace{
      \dfrac{1}{4}\left(
        H_{\dot{a}I\dot{b}J}
      - H_{\dot{b}I\dot{a}J}
      + H_{\dot{b}J\dot{a}I} 
      - H_{\dot{a}J\dot{b}I}
      \right)}_{\displaystyle = H_{[\dot{a}\dot{b}][IJ]} = H_R(6,1,1)_{422}}
    \;,
\cr
& &
\end{eqnarray}
where the parentheses and brackets on the indices respectively indicate 
symmetrization and anti-symmetrization. 
Let us look at the couplings of the two terms separately.
First, $\Delta_{(\dot{a}\dot{b})(IJ)} = \Delta_R(10,1,3)_{422}$:
\begin{eqnarray}
\lefteqn{\overline{\psi^C}_{\dot{a}I\vphantom{\dot{b}}}\gamma_5 \Delta^{(\dot{a}\dot{b})(IJ)} \psi_{\dot{b}J}
\;+\; h.c. \vphantom{\bigg|}} 
\cr 
& = & 
\left(
 \overline{\psi^C}_{\dot{a}0\vphantom{\dot{b}}} \Delta^{(\dot{a}\dot{b})(00)} \psi_{\dot{b}0}
+\overline{\psi^C}_{\dot{a}0\vphantom{\dot{b}}} \Delta^{(\dot{a}\dot{b})(0j)} \psi_{\dot{b}j}
+\overline{\psi^C}_{\dot{a}i\vphantom{\dot{b}}} \Delta^{(\dot{a}\dot{b})(i0)} \psi_{\dot{b}0}
+\overline{\psi^C}_{\dot{a}i\vphantom{\dot{b}}} \Delta^{(\dot{a}\dot{b})(ij)} \psi_{\dot{b}j}
\right)
\;+\; h.c.
\vphantom{\bigg|}
\cr
& = & 
\biggl[
 \Bigl(\overline{L_R^C}\,\epsilon\vec{\tau} L_R^{\vphantom{C}}\Bigr) \vec{\Delta}_{R1}^*
+2\Bigl(\overline{Q_R^C}\,\epsilon\vec{\tau} L_R^{\vphantom{C}}\Bigr) \vec{\Delta}_{R3}^* 
+\Bigl(\overline{Q_R^C}\,\epsilon\vec{\tau} Q_R^{\vphantom{C}}\Bigr) \vec{\Delta}_{R6}^*
\biggr]
\;+\; h.c.
\label{DeltaRcouplings1}
\end{eqnarray}
%
where we use the shorthand
\begin{equation}
\begin{array}{llll}
\vec{\Delta}_{R1} 
& = \; \Delta_{(\dot{a}\dot{b})(00)}
& = \; \Delta_{R1}\left(1,1,3,-2\right)_{3221} & = \;
\left[\begin{array}{l}
\Delta_{R1}(1,1,\phantom{-}0)_{321} \\ \Delta_{R1}(1,1,-1)_{321} \\ \Delta_{R1}(1,1,-2)_{321}
\end{array}\right]
,\\
\vec{\Delta}_{R3} 
& = \; \Delta_{(\dot{a}\dot{b})(0j)}
& = \; \Delta_{R3}\left(3,1,3,-\frac{2}{3}\right)_{3221} & = \;
\left[\begin{array}{l}
\Delta_{R3}\left(3,1,\phantom{-}\frac{2}{3}\right)_{321} \\ 
\Delta_{R3}\left(3,1,-\frac{1}{3}\right)_{321} \\ 
\Delta_{R3}\left(3,1,-\frac{4}{3}\right)_{321}
\end{array}\right]
,\\
\vec{\Delta}_{R6} 
& = \; \Delta_{(\dot{a}\dot{b})(ij)}
& = \; \Delta_{R6}\left(6,1,3,+\frac{2}{3}\right)_{3221} & = \;
\left[\begin{array}{l}
\Delta_{R6}\left(6,1,\phantom{-}\frac{4}{3}\right)_{321} \\ 
\Delta_{R6}\left(6,1,\phantom{-}\frac{1}{3}\right)_{321} \\ 
\Delta_{R6}\left(6,1,-\frac{2}{3}\right)_{321}
\end{array}\right]
.
\end{array}
\end{equation}
The leptoquark fields are $\vec{\Delta}_{R3}$.
Expanding the interaction, we find
\begin{eqnarray}
\lefteqn{
2\Bigl(\overline{Q_R^C}\,\epsilon\vec{\tau} L_R^{\vphantom{C}}\Bigr) \vec{\Delta}_{R3}^*  
\;+\; h.c.
}
\cr
& = & \bigg[
 \sqrt{2}\Bigl(
   \overline{u_R^C}\nu_R^{\vphantom{\dagger}}
 \Bigr)
 \underbrace{\Delta_{R3}^{*(-2/3)}}_{\displaystyle \overline{S}_1^{(-2/3)}}
+\Bigl(
   \overline{u_R^C}e_R^{\vphantom{\dagger}} 
+  \overline{d_R^C}\nu_R^{\vphantom{\dagger}}
 \Bigr)
 \underbrace{\Delta_{R3}^{*(1/3)}}_{\displaystyle S_1^{(1/3)}} 
+\sqrt{2}\Bigl(
   \overline{d_R^C}e_R^{\vphantom{\dagger}}   
 \Bigr) 
 \underbrace{\Delta_{R3}^{*(4/3)}}_{\displaystyle \widetilde{S}_1^{(4/3)}}
\biggr] \;+\; h.c.
\cr
& &
\label{DeltaRcouplings2}
\end{eqnarray}
Thus, the field $\Delta_R(10,1,3)_{422}$ includes components corresponding to the leptoquarks
$S_1$, $\widetilde{S}_1$, and $\overline{S}_1$.
They all couple only to right-handed fermions by construction, and
none of them couple to diquarks.

Next, let us consider the couplings of $H_{[\dot{a}\dot{b}][IJ]} = H_R(6,1,1)_{422}$.
We find:
\begin{eqnarray}
\lefteqn{\overline{\psi^C}_{\dot{a}I\vphantom{\dot{b}}}\gamma_5 H^{[\dot{a}\dot{b}][IJ]} \psi_{\dot{b}J}
\;+\; h.c. 
\vphantom{\bigg|}
} 
\cr 
& = & \left[
 \left(\overline{\psi^C}_{\dot{a}0\vphantom{\dot{b}}}\psi_{\dot{b}j}  
-\overline{\psi^C}_{\dot{a}j\vphantom{\dot{b}}}\psi_{\dot{b}0} \right) H^{[\dot{a}\dot{b}][0j]} 
+\overline{\psi^C}_{\dot{a}i\vphantom{\dot{b}}}\psi_{\dot{b}j} H^{[\dot{a}\dot{b}][ij]} 
\right] \;+\; h.c. 
\vphantom{\bigg|} 
\cr
& = & \biggl[
\left(
 \overline{L_R^C}\epsilon\, Q_{Rj}^{\vphantom{C}}
-\overline{Q_{Rj}^C}\epsilon\, L_R^{\vphantom{C}}
\right) H^{[\dot{1}\dot{2}][0j]}
+ 
\left(
 \overline{Q_{Ri}^C}\epsilon\, Q_{Rj}^{\vphantom{C}}
\right) H^{[\dot{1}\dot{2}][ij]}   
\biggr] 
\;+\; h.c.
\vphantom{\Bigg|} 
\cr
& = & 2\biggl[
 \left(
 \overline{d_{Rj}^C}\nu_R^{\vphantom{\dagger}}
-\overline{u_{Rj}^C}e_R^{\vphantom{\dagger}}  
 \right) \underbrace{H_{3R}^{*j}}_{\displaystyle S_1^{(1/3)}} 
+\,
 \varepsilon^{ijk}
 \overline{u_{Ri}^C}d_{Rj}^{\vphantom{\dagger}} 
 H_{\overline{3}Rk}^{*} 
\biggr] \;+\; h.c. 
\vphantom{\bigg|} 
\label{HRcouplings}
\end{eqnarray}
where in the last line we have set
\begin{eqnarray}
H_{3Rj} & \;=\; & H_{[\dot{1}\dot{2}][0,j]}\;,\cr
H_{\overline{3}R}{}^k & \;=\; & \dfrac{1}{2}\varepsilon^{ijk} H_{[\dot{1}\dot{2}][i,j]}
\qquad\to\qquad
H_{[\dot{1}\dot{2}][i,j]}
\;=\; \varepsilon_{ijk} H_{\overline{3}R}{}^k
\;.
\end{eqnarray}
%
%
%
So $H_{3R}\!\left(3,1,-\frac{1}{3}\right)_{321}$ corresponds to leptoquark 
$S_1^*$ coupled to right-handed fermions.
The field $H_{\overline{3}R}\!\left(\overline{3},1,\frac{1}{3}\right)_{321}$ has the
quantum numbers of $S_1$, but it only couples to diquarks.

Now let us look at the other field $\Sigma_{\dot{a}I}^{bJ}$. 
It is also a product representation,
$(\bar 4,2,1) \times (4,1,2) = (1,2,2) + (15,2,2)$. 
The corresponding fields are denoted $\phi(1,2,2)_{422}$ and $\widetilde\Sigma(15,2,2)_{422}$.
The couplings of $\phi_{\dot{a}}^{b} = \phi(1,2,2)_{422}$ to the fermions are simply
\begin{equation}
\overline{\psi}{}^{\dot{a}I}
\gamma_5 
\left( k^u \phi^{b}_{\dot{a}} + k^d \tilde{\phi}^{b}_{\dot{a}}\right)
\psi_{bI\vphantom{\dot{b}}}^{\vphantom{\dagger}} 
\;+\; h.c.
\end{equation}
where $k^{u,d}$ are Yukawa coupling matrices,
which will give the same masses to the quarks and leptons upon symmetry breaking to $G_{31}$.
This quark-lepton symmetry is broken by the couplings to 
$\widetilde{\Sigma}_{\dot{a}I}^{bJ} = \widetilde{\Sigma}(15,2,2)_{422}$.
They read:
\begin{eqnarray}
\lefteqn{
\Bigl(\overline{\psi}{}^{\dot{a}I}
\gamma_5 \widetilde{\Sigma}^{bJ}_{\dot{a}I\vphantom{\dot{b}}} 
\psi_{bJ\vphantom{\dot{b}}}^{\vphantom{\dagger}}
\Bigr) 
\;+\; h.c.}
\cr
& = &
-\Big(
 \overline{\psi}{}^{\dot{a}0}\,\widetilde{\Sigma}_{\dot{a}0}^{b0}\,\psi_{b0}^{\vphantom{\dagger}}
+\overline{\psi}{}^{\dot{a}i}\,\widetilde{\Sigma}_{\dot{a}i}^{b0}\,\psi_{b0}^{\vphantom{\dagger}}
+\overline{\psi}{}^{\dot{a}0}\,\widetilde{\Sigma}_{\dot{a}0}^{bj}\,\psi_{bj}^{\vphantom{\dagger}}
+\overline{\psi}{}^{\dot{a}i}\,\widetilde{\Sigma}_{\dot{a}i}^{bj}\,\psi_{bj}^{\vphantom{\dagger}}
\Bigr) \;+\; h.c.
\cr
& = &
-\Bigl(
 \overline{L_R}\widetilde{\Sigma}^\dagger_1 L_L
+\overline{Q_R}\widetilde{\Sigma}^\dagger_{\overline{3}} L_L
+\overline{L_R}\widetilde{\Sigma}^\dagger_{3} Q_L
+\overline{Q_R}\widetilde{\Sigma}^\dagger_{8} Q_L
\Bigr) \;+\; h.c.
\end{eqnarray}
where
\begin{equation}
\begin{array}{llll}
\widetilde{\Sigma}_{1}
& = \;
\widetilde{\Sigma}_{1}\left(1,2,2,0\right)_{3221}
& = \;
\left[
\begin{array}{ll}
\widetilde{\Sigma}'_{1}\left(1,2,-\frac{1}{2}\right)_{321} &
\widetilde{\Sigma}_{1} \left(1,2,\phantom{-}\frac{1}{2}\right)_{321}
\end{array}
\right]
& = \;
\left[
\begin{array}{ll}
\widetilde{\Sigma}_{1}^{\prime (0)}  & \widetilde{\Sigma}_{1}^{(1)} \\
\widetilde{\Sigma}_{1}^{\prime (-1)} & \widetilde{\Sigma}_{1}^{(0)} \\
\end{array}
\right]
\;,
\\
\widetilde{\Sigma}_{3}
& = \; 
\widetilde{\Sigma}_{3}\left(3,2,2,+\frac{4}{3}\right)_{3221}
& = \;
\left[
\begin{array}{ll}
\widetilde{\Sigma}'_{3}\left(3,2,+\frac{1}{6}\right)_{321} &
\widetilde{\Sigma}_{3} \left(3,2,+\frac{7}{6}\right)_{321}
\end{array}
\right]
& = \;
\left[
\begin{array}{ll}
\widetilde{\Sigma}_{3}^{\prime (2/3)}  & \widetilde{\Sigma}_{3}^{(5/3)} \\
\widetilde{\Sigma}_{3}^{\prime (-1/3)} & \widetilde{\Sigma}_{3}^{(2/3)} \\
\end{array}
\right]
\;,
\\
\widetilde{\Sigma}_{\overline{3}}
& = \; 
\widetilde{\Sigma}_{\overline{3}}\left(\overline{3},2,2,-\frac{4}{3}\right)_{3221}
& = \;
\left[
\begin{array}{ll}
\widetilde{\Sigma}'_{\overline{3}} \left(\overline{3},2,-\frac{7}{6}\right)_{321} &
\widetilde{\Sigma}_{\overline{3}}\left(\overline{3},2,-\frac{1}{6}\right)_{321}
\end{array}
\right]
& = \;
\left[
\begin{array}{ll}
\widetilde{\Sigma}_{\overline{3}}^{\prime (-2/3)} & \widetilde{\Sigma}_{\overline{3}}^{(1/3)} \\
\widetilde{\Sigma}_{\overline{3}}^{\prime (-5/3)} & \widetilde{\Sigma}_{\overline{3}}^{(-2/3)} \\
\end{array}
\right]
\;,
\\
\widetilde{\Sigma}_{8}
& = \;
\widetilde{\Sigma}_{8}\left(8,2,2,0\right)_{3221}
& = \;
\left[
\begin{array}{ll}
\widetilde{\Sigma}'_{8}\left(8,2,-\frac{1}{2}\right)_{321} &
\widetilde{\Sigma}_{8}\left(8,2,\phantom{-}\frac{1}{2}\right)_{321}
\end{array}
\right]
& = \;
\left[
\begin{array}{ll}
\widetilde{\Sigma}_{8}^{\prime (0)}  & \widetilde{\Sigma}_{8}^{(1)} \\
\widetilde{\Sigma}_{8}^{\prime (-1)} & \widetilde{\Sigma}_{8}^{(0)} \\
\end{array}
\right]
\;.
\end{array}
\end{equation}
Only the uncolored field $\widetilde{\Sigma}_1$ can develop a VEV upon breaking to $G_{31}$,
which will only give masses to the leptons
and break quark-lepton universality.
The leptoquark fields are $\widetilde{\Sigma}_{3}$ and $\widetilde{\Sigma}_{\overline{3}}$:
Expanding out the couplings explicitly, we find:
\begin{eqnarray}
\overline{Q_R}\widetilde{\Sigma}^\dagger_{\overline{3}} L_L
& = & 
\biggl[
 \bigl(\overline{u_R}\nu_L\bigr) 
 \underbrace{\widetilde{\Sigma}_{\overline{3}}^{*(2/3)}}_{\displaystyle R_2^{(2/3)}}
+\bigl(\overline{u_R}e_L  \bigr) 
 \underbrace{\widetilde{\Sigma}_{\overline{3}}^{*(5/3)}}_{\displaystyle R_2^{(5/3)}}
+\bigl(\overline{d_R}\nu_L\bigr) 
 \underbrace{\widetilde{\Sigma}_{\overline{3}}^{\prime *(-1/3)}}_{\displaystyle \widetilde{R}_2^{(-1/3)}}
+\bigl(\overline{d_R}e_L  \bigr) 
 \underbrace{\widetilde{\Sigma}_{\overline{3}}^{\prime *(2/3)}}_{\displaystyle \widetilde{R}_2^{(2/3)}}
\biggr] 
\;,\cr
\overline{L_R}\widetilde{\Sigma}^\dagger_{3} Q_L 
& = & 
\biggl[
 \bigl(\overline{\nu_R}u_L\bigr) 
 \underbrace{\widetilde{\Sigma}_{3}^{\prime *(-2/3)}}_{\displaystyle \widetilde{R}_2^{*(-2/3)}}
+\bigl(\overline{\nu_R}d_L\bigr) 
 \underbrace{\widetilde{\Sigma}_{3}^{\prime *(1/3)}}_{\displaystyle \widetilde{R}_2^{*(1/3)}}
+\bigl(\overline{e_R}u_L  \bigr) 
 \underbrace{\widetilde{\Sigma}_{3}^{*(-5/3)}}_{\displaystyle R_2^{*(-5/3)}}
+\bigl(\overline{e_R}d_L  \bigr) 
 \underbrace{\widetilde{\Sigma}_{3}^{*(-2/3)}}_{\displaystyle R_2^{*(-2/3)}}
\biggr] 
\;.\cr
& &
\end{eqnarray}
Note that though both $\widetilde{\Sigma}_{\overline{3}}^{*(2/3)}$ and
$\widetilde{\Sigma}_{3}^{(2/3)}$, for instance, have the quantum numbers and couplings of the leptoquark $R_2^{(2/3)}$, they are independent fields.
Note also that these leptoquarks do not couple to diquarks.


\subsubsection{Model C}
\label{sec:model-c}

As was discussed above, Model C is the left-right symmetrization of Model B. 
Besides all the fields listed in Model B
that couple to right-handed fermions, it has the counterpart scalars
that couple to left-handed fermions, 
$H_{aIbJ} = \Delta_{(ab)(IJ)} + H_{[ab][IJ]} = \Delta_L(10,3,1)_{422} + H_L(6,1,1)_{422}$, in addition to 
$H_{\dot{a}I\dot{b}J} = \Delta_{(\dot{a}\dot{b})(IJ)} + H_{[\dot{a}\dot{b}][IJ]} = \Delta_R(10,1,3)_{422} + H_R(6,1,1)_{422}$. 
The couplings of $\Delta_L(10,3,1)_{422}$ and $H_L(6,1,1)_{422}$ to the fermions can be obtained 
from those of $\Delta_R(10,1,3)_{422}$ and $H_R(6,1,1)_{422}$,  Eqs.~\eqref{DeltaRcouplings1},
\eqref{DeltaRcouplings2}, and \eqref{HRcouplings}, by simply replacing the right-handed fermions
with their left-handed counterparts.
First, the couplings of $\Delta_L(10,3,1)_{422}$ are
\begin{eqnarray}
\lefteqn{\overline{\psi^C}_{aI}\gamma_5 \Delta^{(ab)(IJ)} \psi_{bJ}^{\vphantom{C}}
\;+\; h.c. \vphantom{\Big|}} 
\cr
& = &
\biggl[
  \Bigl(\overline{L_L^C}\,\epsilon\vec{\tau} L_L^{\vphantom{C}}\Bigr) \vec{\Delta}_{L1}^*
+2\Bigl(\overline{Q_L^C}\,\epsilon\vec{\tau} L_L^{\vphantom{C}}\Bigr) 
  \underbrace{\vec{\Delta}_{L3}^*}_{\displaystyle \vec{S}_3} 
+ \Bigl(\overline{Q_L^C}\,\epsilon\vec{\tau} Q_L^{\vphantom{C}}\Bigr) \vec{\Delta}_{L6}^*
\biggr]
\;+\; h.c.
\label{DeltaLcouplings1}
\end{eqnarray}
Expanding the $\vec{\Delta}_{L3}^*$ term we obtain
\begin{eqnarray}
\lefteqn{
2\Bigl(\overline{Q_L^C}\,\epsilon\vec{\tau} L_L^{\vphantom{C}}\Bigr) \vec{\Delta}_{L3}^*  
\;+\; h.c.
}
\cr
& = & \bigg[
 \sqrt{2}\Bigl(
   \overline{u_L^C}\nu_L^{\vphantom{\dagger}}
 \Bigr)
 \underbrace{\Delta_{L3}^{*(-2/3)}}_{\displaystyle S_3^{(-2/3)}}
+\Bigl(
   \overline{u_L^C}e_L^{\vphantom{\dagger}} 
+  \overline{d_L^C}\nu_L^{\vphantom{\dagger}}
 \Bigr)
 \underbrace{\Delta_{L3}^{*(1/3)}}_{\displaystyle S_3^{(1/3)}} 
+\sqrt{2}\Bigl(
   \overline{d_L^C}e_L^{\vphantom{\dagger}}   
 \Bigr) 
 \underbrace{\Delta_{L3}^{*(4/3)}}_{\displaystyle S_3^{(4/3)}}
\biggr] \;+\; h.c.
\cr
& &
\label{DeltaLcouplings2}
\end{eqnarray}
Thus, we obtain the isotriplet leptoquark $S_3$.
The couplings of $H_L(6,1,1)_{422}$ on the other hand are
\begin{eqnarray}
\lefteqn{
\overline{\psi^C}_{aI}\gamma_5 H^{[ab][IJ]} \psi_{bJ}^{\vphantom{C}} \;+\; h.c.
\vphantom{\Big|}
}\cr
& = & 2\biggl[
\Bigl(\overline{d_{Lj}^C}\nu_L^{\vphantom{\dagger}} - \overline{u_{Lj}^C}e_L^{\vphantom{\dagger}}\Bigr)
\underbrace{H_{3L}^{*j}}_{\displaystyle S_1^{(1/3)}} 
+\;
\varepsilon^{ijk}\overline{u_{Li}^C}d_{Lj}^{\phantom{\dagger}}H_{\overline{3}Lk}^{*}
\biggr]
\;+\; h.c.
\label{HLcouplings}
\end{eqnarray}
and we finally find the leptoquark we seek:
a leptoquark $S_1 = H_{3L}^*\!\left(\overline{3},1,\frac{1}{3}\right)_{321}$ which couples to left-handed fermions.
The field $H_{\overline{3}L}\!\left(\overline{3},1,\frac{1}{3}\right)_{321}$ has the same quantum numbers, but this one couples only to diquarks.
Thus, we have surveyed the three NCG models and found a unique field, $H_{3L}\!\left(3,1,-\frac{1}{3}\right)_{321}$
in Model C, which serves the purpose of the leptoquark 
$S_1\!\left(\overline{3},1,\frac{1}{3}\right)_{321}$ coupled to left-handed fermions.



\newpage

\begin{table}[!h]
\vspace{-1cm}
{\small
\begin{center}
\begin{tabular}{l|l|l}
\hline
$G_{422}$ & $G_{3221}$ & $G_{321}$ \\ 
\hline\hline
$\phi(1,2,2)_{422}$ 
& $\phi(1,2,2,0)_{3221}$ 
& $\phi_2 \!\left(1,2, \frac{1}{2}\right)_{321}$,
  $\phi_2'\!\left(1,2,-\frac{1}{2}\right)_{321}$ 
\vphantom{\Bigg|}\\
\hline
$\widetilde{\Delta}_R\left(4,1,2\right)_{422}$
& $\widetilde{\Delta}_{R1}\!\left(1,1,2,-1\right)_{3221}$ 
& $\widetilde{\Delta}_{R1}\left(1,1,0\right)_{321}$,
  $\widetilde{\Delta}_{R1}\left(1,1,-1\right)_{321}$ 
\vphantom{\Bigg|}\\
\cline{2-3}
& $\widetilde{\Delta}_{R3}\!\left(3,1,2,\frac{1}{3}\right)_{3221}$ 
& $\widetilde{\Delta}_{R3}\!\left(3,1,\frac{2}{3}\right)_{321}$, 
  $\widetilde{\Delta}_{R3}\!\left(3,1,-\frac{1}{3}\right)_{321}$ 
\vphantom{\Bigg|}\\
\hline
$\Delta_R\left(10,1,3\right)_{422}$
& $\Delta_{R1}\left(1,1,3,-2\right)_{3221}$
& $\Delta_{R1}\left(1,1,0\right)_{321}$,
  $\Delta_{R1}\left(1,1,-1\right)_{321}$,
  $\Delta_{R1}\left(1,1,-2\right)_{321}$ 
\vphantom{\bigg|}\\
\cline{2-3}
& $\Delta_{R3}\!\left(3,1,3,-\frac{2}{3}\right)_{3221}$
& $\Delta_{R3}\!\left(3,1,\frac{2}{3}\right)_{321}$,
  $\Delta_{R3}\!\left(3,1,-\frac{1}{3}\right)_{321}$,
  $\Delta_{R3}\!\left(3,1,-\frac{4}{3}\right)_{321}$ 
\vphantom{\Bigg|}\\
\cline{2-3}
& $\Delta_{R6}\!\left(6,1,3,\frac{2}{3}\right)_{3221}$
& $\Delta_{R6}\!\left(6,1,\frac{4}{3}\right)_{321}$,
  $\Delta_{R6}\!\left(6,1,\frac{1}{3}\right)_{321}$,
  $\Delta_{R6}\!\left(6,1,-\frac{2}{3}\right)_{321}$ 
\vphantom{\Bigg|}\\
\hline
$\Delta_L\left(10,3,1\right)_{422}$
& $\Delta_{L1}\left(1,3,1,-2\right)_{3221}$
& $\Delta_{L1}\left(1,3,-1\right)_{321}$ 
\vphantom{\bigg|}\\
\cline{2-3}
& $\Delta_{L3}\!\left(3,3,1,-\frac{2}{3}\right)_{3221}$
& $\Delta_{L3}\!\left(3,3,-\frac{1}{3}\right)_{321}$ 
\vphantom{\Bigg|}\\
\cline{2-3}
& $\Delta_{L6}\!\left(6,3,1,\frac{2}{3}\right)_{3221}$
& $\Delta_{L6}\!\left(6,3,\frac{1}{3}\right)_{321}$ 
\vphantom{\Bigg|}\\
\hline
$H_{R/L}(6,1,1)_{422}$
& $H_{3R/L}\!\left(3,1,1,-\frac{2}{3}\right)_{3221}$
& $H_{3R/L}\!\left(3,1,-\frac{1}{3}\right)_{321}$ 
\vphantom{\Bigg|} \\
\cline{2-3}
& $H_{\bar{3}R/L}\!\left(\bar{3},1,1,\frac{2}{3}\right)_{3221}$
& $H_{\bar{3}R/L}\!\left(\bar{3},1,\frac{1}{3}\right)_{321}$ 
\vphantom{\Bigg|} \\
\hline
$\Sigma\left(15,1,1\right)_{422}$ 
& $\Sigma_1\left(1,1,1,0\right)_{3221}$ 
& $\Sigma_1\left(1,1,0\right)_{321}$ 
\vphantom{\bigg|}\\
\cline{2-3}
& $\Sigma_3\!\left(3,1,1,\frac{4}{3}\right)_{3221}$ 
& $\Sigma_3\!\left(3,1,\frac{2}{3}\right)_{321}$ 
\vphantom{\Bigg|}\\ 
\cline{2-3}
& $\Sigma_{\bar{3}}\!\left(\bar{3},1,1,-\frac{4}{3}\right)_{3221}$
& $\Sigma_{\bar{3}}\!\left(\bar{3},1,-\frac{2}{3}\right)_{321}$ 
\vphantom{\Bigg|}\\
\cline{2-3}
& $\Sigma_8\left(8,1,1,0\right)_{3221}$ 
& $\Sigma_8\left(8,1,0\right)_{321}$
\vphantom{\bigg|}\\ 
\hline
$\widetilde{\Sigma}\left(15,2,2\right)_{422}$
& $\widetilde{\Sigma}_1\left(1,2,2,0\right)_{3221}$
& $\widetilde{\Sigma}_1 \!\left(1,2, \frac{1}{2}\right)_{321}$, 
  $\widetilde{\Sigma}_1'\!\left(1,2,-\frac{1}{2}\right)_{321}$ 
\vphantom{\Bigg|}\\
\cline{2-3}
& $\widetilde{\Sigma}_3 \!\left(3,2,2,\frac{4}{3}\right)_{3221}$ 
& $\widetilde{\Sigma}_3 \!\left(3,2,\frac{7}{6}\right)_{321}$,
  $\widetilde{\Sigma}_3'\!\left(3,2,\frac{1}{6}\right)_{321}$ 
\vphantom{\Bigg|}\\
\cline{2-3}
& $\widetilde{\Sigma}_{\bar{3}} \!\left(\bar{3},2,2,-\frac{4}{3}\right)_{3221}$ 
& $\widetilde{\Sigma}_{\bar{3}} \!\left(\bar{3},2,-\frac{1}{6}\right)_{321}$,
  $\widetilde{\Sigma}_{\bar{3}}'\!\left(\bar{3},2,-\frac{7}{6}\right)_{321}$ 
\vphantom{\Bigg|}\\
\cline{2-3}
& $\widetilde{\Sigma}_8\left(8,2,2,0\right)_{3221}$ 
& $\widetilde{\Sigma}_8 \!\left(8,2, \frac{1}{2}\right)_{321}$,
  $\widetilde{\Sigma}_8'\!\left(8,2,-\frac{1}{2}\right)_{321}$
\vphantom{\Bigg|}\\ 
\hline
\end{tabular}
\end{center}
}
\caption{\label{Decompositions}
The decomposition of various $G_{422}$ representations into those of $G_{3221}$ and $G_{321}$ (SM). 
The table is from Ref.~\cite{Aydemir:2016xtj} except  
signs and normalizations of the charges have been changed to
conform to the more commonly used convention so that
$Q_{\rm em}=I_{3}^L+Y=I_{3}^L+I_{3}^R+(B-L)/2$. 
All the color triplet fields have quantum numbers corresponding to
some leptoquark though that does not guarantee that they couple to
lepton-quark pairs.
}
\end{table}

\newpage
\section{Gauge Coupling Unification}

In the following, we inquire whether the required unification of gauge couplings in the NCG models can be achieved. We focus specifically on Model C since it is the only model, among the NCG based Pati-Salam models, that contains the required leptoquark  $H_{3L}\left(3,1,-\frac{1}{3}\right)_{321}$.  
We find that in Model C with a light leptoquark $H_{3L}\left(3,1,-\frac{1}{3}\right)_{321}$, it is necessary for at least one intermediate symmetry breaking scale to exist between the unification scale $M_U$, the scale at which the model emerges 
from an underlying NCG with a unified coupling, 
and the electroweak scale $M_Z$. 
The gauge couplings in these models do not unify otherwise. 
We consider several versions of the model with a single intermediate scale $M_C$
($M_Z<M_C<M_U$) as illustrative examples.
There, the symmetry $G_{422D}$ that Model C possesses when it emerges at $M_U$ 
is assumed to persist down to $M_C$, 
at which point it breaks to the $G_{321}$ of the SM. Note that this specific sequence of symmetry breaking appears to be favoured by various phenomenological bounds~\cite{Altarelli:2013aqa}.

\subsection {1-loop renormalization group running}

For a given particle content, 
the gauge couplings evolve under 1-loop renormalization group (RG) running
in an energy interval $\left[M_A,M_B\right]$ following
\begin{eqnarray}
\label{1looprunning}
\frac{1}{g_{i}^{2}(M_A)} - \dfrac{1}{g_{i}^2(M_B)}
\;=\; \dfrac{a_i}{8 \pi^2}\ln\dfrac{M_B}{M_A}
\;,
\end{eqnarray}
where the RG coefficients $a_i$ are given by \cite{Jones:1981we,Lindner:1996tf}
\begin{eqnarray}
\label{1loopgeneral}
a_{i}
\;=\; -\frac{11}{3}C_{2}(G_i)
& + & \frac{2}{3}\sum_{R_f} T_i(R_f)\cdot d_1(R_f)\cdots d_n(R_f) \cr
& + & \frac{\eta}{3}\sum_{R_s} T_i(R_s)\cdot d_1(R_s)\cdots d_n(R_s)\;.
\end{eqnarray}
The summations in Eq.~(\ref{1loopgeneral}) are, respectively, over irreducible chiral representations of fermions ($R_f$) in the second term and those of scalars ($R_s$) in the third. The coefficient $\eta$ is either 1 or 1/2, depending on whether the corresponding representation is complex or (pseudo) real, respectively. 
%
%
$C_2(G_i)$ is the quadratic Casimir for the adjoint representation of the group $G_i$,
and $T_i$ is the Dynkin index of each representation. 
For the $U(1)$ group, $C_2(G)=0$ and
\begin{equation}
\sum_{f,s}T \;=\; \sum_{f,s}Y^2\;,
\label{U1Dynkin}
\end{equation}
where $Y$ is the $U(1)$ charge.
Particles that are heavier than $M_B$ (we assume $M_A<M_B$) decouple from the running and do not contribute
to the sums in Eq.~(\ref{1loopgeneral}).
So the values of the RG coefficients $a_i$ change every time a symmetry breaking scale
is crossed and some of the particles acquire masses of the order of that scale.



The low energy data which we will use as the boundary conditions to the RG running (in the $\overline{\mathrm{MS}}$ scheme) are
\cite{Patrignani:2016xqp,ALEPH:2005ab}
\begin{eqnarray}
\alpha^{-1}(M_Z) & = & 127.950\pm0.017\;,\cr
\alpha_s(M_Z) & = & 0.1182\pm0.0016\;,\cr
\sin^2\theta_W(M_Z) & = & 0.23129\pm0.00005\;,
\label{SMboundary}
\end{eqnarray}
at $M_Z=91.1876\pm 0.0021\,\mathrm{GeV}$, which translate to
\begin{eqnarray}
g_1(M_Z) & = & 0.357419\pm0.000026\;,\cr
g_2(M_Z) & = & 0.651765\pm0.000083\;,\cr
g_3(M_Z) & = & 1.21875\pm0.00825\;.
\label{MZboundary}
\end{eqnarray}
Throughout the RG running from the unification scale $M_U$ down to the electroweak scale $M_Z$, the coupling constants are all required to remain in the perturbative regime to
justify the use of Eq.~(\ref{1looprunning}).

\begin{table}[t!]
\begin{center}
{\begin{tabular}{ccccc}
\hline
\ \ \ Representation\ \ \ \ \ & $\qquad SU(2)\qquad$ & $\qquad SU(3)\qquad$ & $\qquad SU(4)\qquad$ \\ 
\hline\hline
$\vphantom{\bigg|}$ 2 &   $\dfrac{1}{2}$ &              $-$ &   $-$ & $\vphantom{\bigg|}$ \\
$\vphantom{\bigg|}$ 3 &                2 &   $\dfrac{1}{2}$ &   $-$ & $\vphantom{\bigg|}$ \\
$\vphantom{\bigg|}$ 4 &                5 &              $-$ &   $\dfrac{1}{2}$ & $\phantom{\bigg|}$ \\ 
$\vphantom{\bigg|}$ 6 &  $\dfrac{35}{2}$ &   $\dfrac{5}{2}$ &   $1$ & $\vphantom{\bigg|}$ \\
$\vphantom{\bigg|}$ 8 &               42 &              $3$ &   $-$ & $\vphantom{\bigg|}$ \\
$\vphantom{\bigg|}$10 & $\dfrac{165}{2}$ &  $\dfrac{15}{2}$ &   $3$ & $\vphantom{\bigg|}$ \\
$\vphantom{\bigg|}$15 &              280 & $10,\dfrac{35}{2}$ &  4 & $\vphantom{\bigg|}$ \\
\hline
\end{tabular}
\caption{Dynkin index $T_i$ for several irreducible representations of $SU(2)$, $SU(3)$, and $SU(4)$. 
Different normalization conventions are used in the literature. 
For instance, there is a factor of 2 difference between those given in Ref.~\cite{Lindner:1996tf} and those in Ref.~\cite{Slansky:1981yr}. Our convention follows the former.
Notice that for $SU(3)$, there exist two inequivalent 15 dimensional irreducible representations.
}
\label{DynkinIndex}}
\end{center}
\end{table}

\subsection{Does the leptoquark help coupling unification?}

We first illustrate that unification of the couplings cannot be realized when no intermediate
symmetry breaking scales between $M_U$ and $M_Z$ are present.
Since the symmetry just above $M_Z$ must be that of the SM, $G_{321}$, we are considering the
situation in which the symmetry of Model C breaks immediately to 
$G_{321}$ at $M_U$, the scale at which the model emerges with a higher symmetry 
$G_{422D}$ from the underlying NCG.

Thus the symmetry of the models in the interval $I=[M_Z,M_U]$ is $G_{321}$, and
the vector and fermion content is just that of the SM.
For the scalar content below $M_U$, we consider the fields coming from $\phi(1,2,2)_{422}$ and
$H_L(6,1,1)_{422}$:
\begin{equation}
\begin{array}{lll}
\phi(1,2,2)_{422} 
& \quad\to\quad \phi_2\!\left(1,2,\frac{1}{2}\right)_{321} & + \; \phi_2'\!\left(1,2,-\frac{1}{2}\right)_{321}
\;, \vphantom{\Big|}\\
H_L(6,1,1)_{422}
& \quad\to\quad H_{3L}\!\left(3,1,-\frac{1}{3}\right)_{321} & + \; H_{\overline{3}L}\!\left(\overline{3},1,\frac{1}{3}\right)_{321}
\;. \vphantom{\Big|}
\end{array}
\end{equation}
We identify $\phi_2\left(1,2,\frac{1}{2}\right)_{321}$ with the SM doublet,  
and $H_{3L}\!\left(3,1,-\frac{1}{3}\right)_{321}$ is the scalar leptoquark necessary to explain the $B$-decay anomalies.
We assume that the $H_{\overline{3}L}\left(\overline{3},1,\frac{1}{3}\right)_{321}$ field,
which only couples to diquarks, 
becomes heavy when $SU(4)$ breaks to $SU(3)$ and does not survive to low energies.
For the second Higgs doublet $\phi'_2\left(1,2,-\frac{1}{2}\right)_{321}$,
we have a choice of allowing it to survive to low energies after the breaking of $SU(2)_R$, 
in which case we have a 2-Higgs doublet model (2HDM), or making it heavy and decoupling it after the breaking.

As the scalar content, 
we will therefore consider the following two cases:
\begin{equation}
\label{model0}
\begin{array}{ll}
I_1 &\;:\;\; \phi_2\!\left(1,2,\frac{1}{2}\right)_{321}\mbox{the SM Higgs},\;\; 
H_{3L}\!\left(3,1,-\frac{1}{3}\right)_{321}
\vphantom{\Big|}\\
I_2 &\;:\;\; 
\phi_2\!\left(1,2,\frac{1}{2}\right)_{321}\mbox{the SM Higgs},\;\; 
\phi'_2\!\left(1,2,-\frac{1}{2}\right)_{321}
,\;\; 
H_{3L}\!\left(3,1,-\frac{1}{3}\right)_{321}
\vphantom{\Big|}
\end{array}
\end{equation}
The rest of the degrees of freedom are assumed to become heavy at the unification scale $M_U$.
%
With this particle content, the RG coefficients are given by
$a_i \;=\; a_i^\mathrm{SM}+\Delta a_i\;$,  $(i=1,2,3)$
where
\begin{eqnarray}
a_i^\mathrm{SM} & = & \left[\,\dfrac{41}{6}\,,\;-\dfrac{19}{6}\,,\;-7\,\right]\;, \cr
\Delta a_i & = & \left[\,\dfrac{1}{9}\left(\dfrac{5}{18}\right)\,,\;0\left(\dfrac{1}{6}\right)\,,\;\dfrac{1}{6}\left(\dfrac{1}{6}\right)\right]\;,
\label{RGcoefI12}
\end{eqnarray}
for model $I_1$ ($I_2$).

The symmetry of the model $G_{442D}$ is broken to the symmetry of the SM $G_{321}$ at $M_U$, which is also the scale where it emerges from an underlying NCG. 
The obvious boundary/matching conditions to be imposed on the couplings at $M_U$ and $M_Z$ 
are:
\begin{eqnarray}
M_U & \;:\; & \sqrt{\frac{5}{3}}\;g_1(M_U) \;=\; g_2(M_U) \;=\; g_3(M_U) \;, \vphantom{\bigg|} 
\cr
M_Z & \;:\; & \frac{1}{e^2(M_Z)} \;=\; \frac{1}{g_1^2(M_Z)}+\frac{1}{g_2^2(M_Z)}\;.\vphantom{\Bigg|} 
\label{Matching0}
\end{eqnarray}
Using Eq.~\eqref{1looprunning} with the low energy data and boundary conditions given in Eqs.~\eqref{SMboundary} and \eqref{Matching0}, the following equations are obtained.
\begin{eqnarray}
\label{equation0}
2\pi\left[\frac{3-8 \sin^2\theta_w (M_Z)}{\alpha(M_Z)}\right]&=&(3 a_1-5 a_2)\ln\frac{M_U}{M_Z} \;,\cr
2\pi\left[\frac{3}{\alpha(M_Z)}-\frac{8}{\alpha_s (M_Z)}\right]&=& (3 a_1+3 a_2-8 a_3)\ln\frac{M_U}{M_Z}\;.
\end{eqnarray}
Taking the ratio of these equations, which also cancels the uncertainty on $M_Z$, yields the condition to be satisfied for unification as
\begin{equation}\label{constraint}
r\;\equiv\;\frac{3a_1-5a_2}{3a_1+3a_2-8a_3}\;\cong\; 0.4656\pm0.0014\;.
\end{equation}
Using Eq.~\eqref{RGcoefI12} we find that
the models $I_1$ and $I_2$ yield $r_1\cong0.56$ and $r_2\cong0.54$, respectively. 
Neither of these values fall in the required range stated above. 
The latter value, $r_2$, is actually identical to $r_\mathrm{SM}$ by coincidence; the modifications in both the numerator and denominator of $r_2$ cancel. 
Thus, allowing the second Higgs doublet $\phi_2'\left(1,2,\frac{1}{2}\right)_{321}$ to survive in model $I_2$
does not help in unifying the couplings.

\begin{figure}[t]
\subfigure[ ]{\includegraphics[width=7cm]{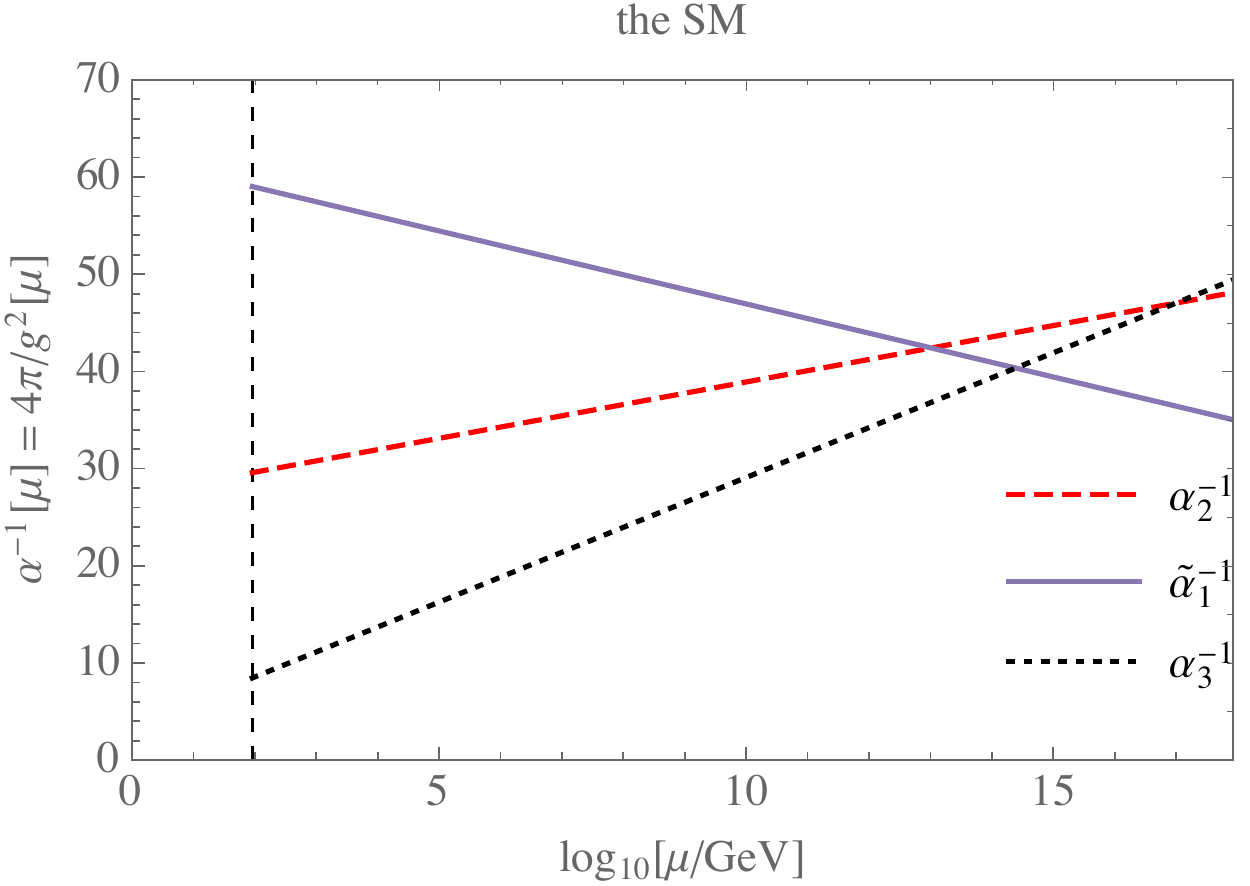}}\hspace{0.4cm}
\subfigure[ ]{\includegraphics[width=7cm]{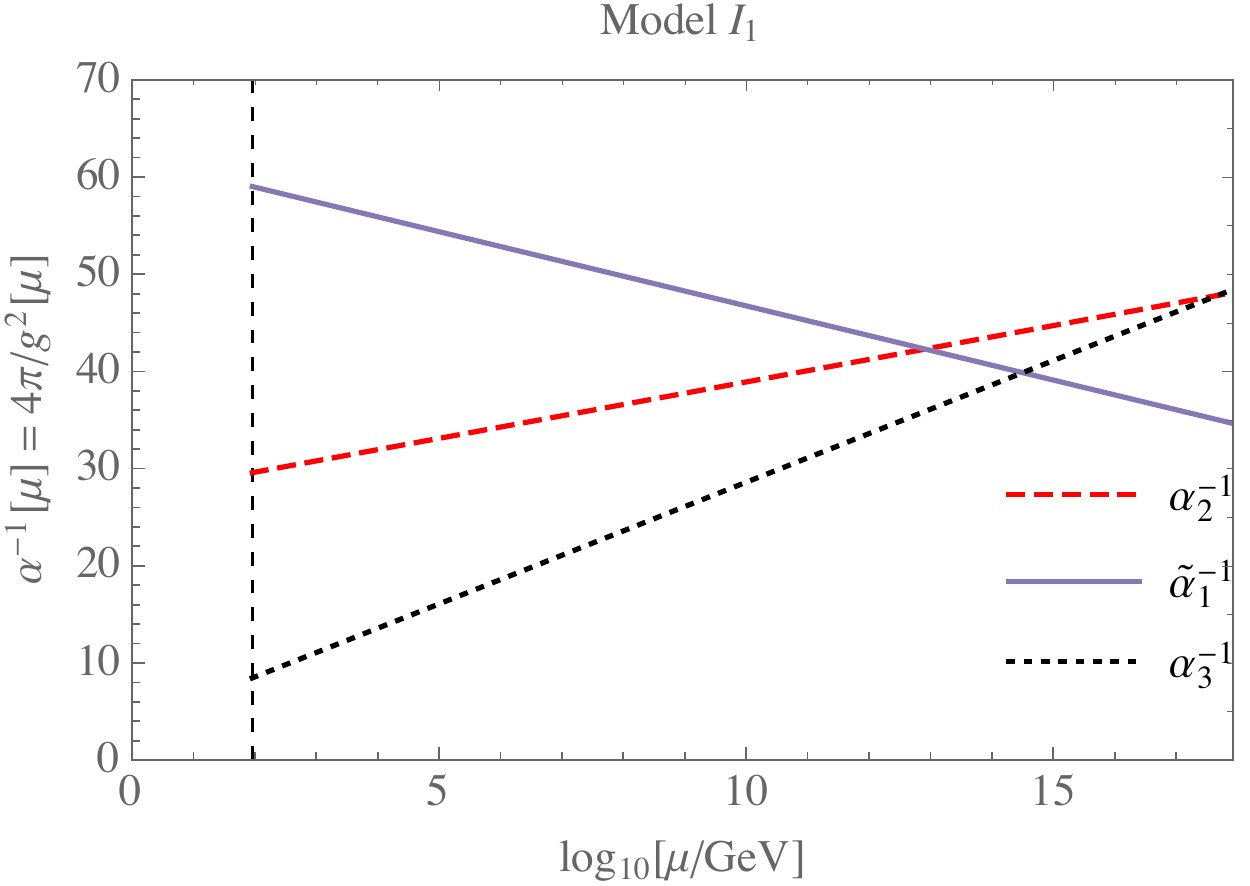}}\vspace{0.4cm}
\subfigure[ ]{\includegraphics[width=7cm]{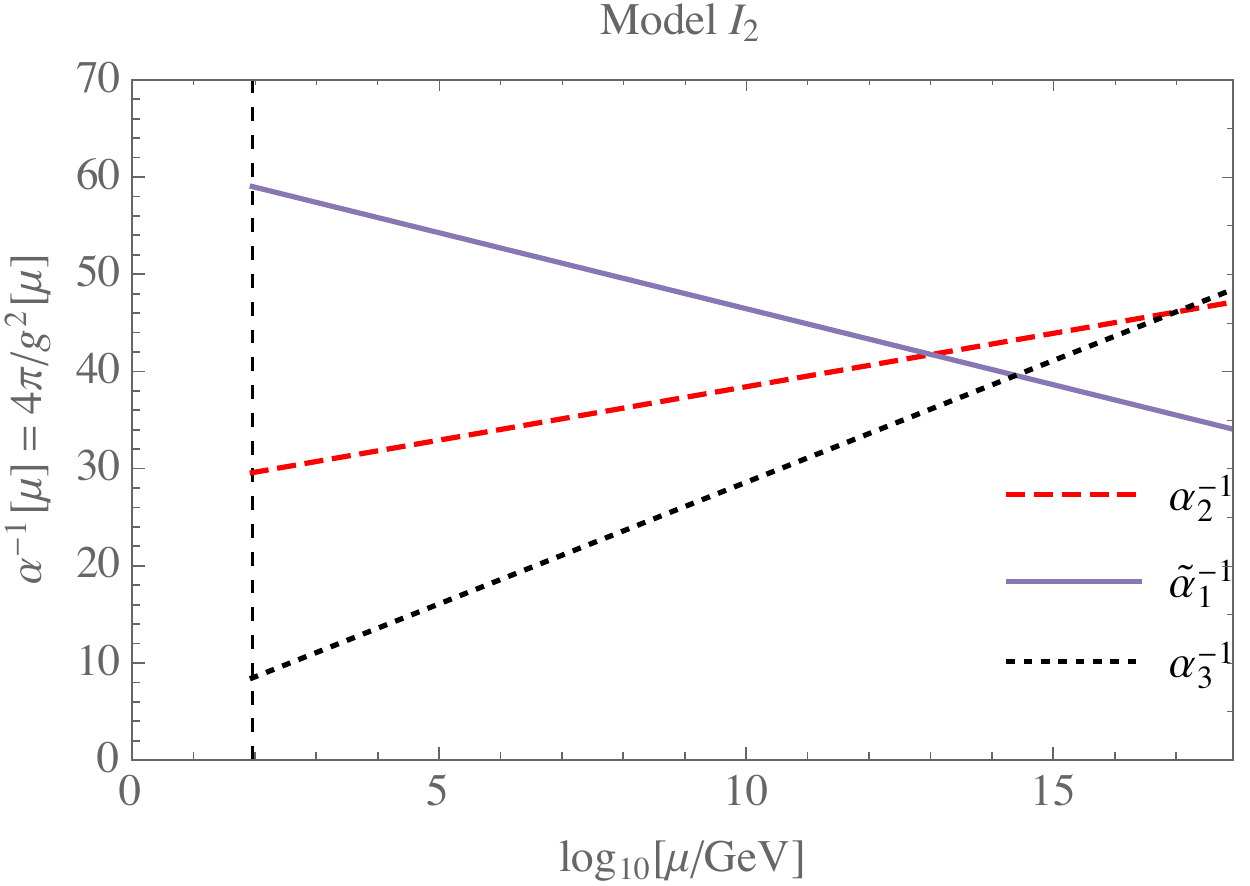}}\hspace{0.4cm}
\caption{Running of the gauge couplings for the SM, Model $I_1$, and Model $I_2$. The vertical dotted line correspond to the electroweak scale $M_Z$. For $\alpha_1^{-1}$, we plot the redefined quantity $\widetilde{\alpha}^{-1}_1\equiv \frac{3}{5}\alpha^{-1}_1$. Including the leptoquark with or without a second Higgs doublet does not make a significant modification to the SM running in favor of unification. Therefore in the NCG framework, one needs to consider intermediate symmetry breaking scales for these options. Note that the running of the couplings in the case of the SM and Model $I_2$ is almost identical due to accidental cancellations of additional contributions in the 1-loop RG coefficients ($\Delta a_i$) of Model $I_2$.
}
\label{RGrunning1}
\end{figure}

The resulting non-unification of the couplings can also be observed in Figure~\ref{RGrunning1}, where their RG runnings are displayed. Notice that the overall running of the couplings in Model $I_1$ and Model $I_2$ are almost identical to the one in the SM due to the smallness of the ratios $\Delta a_i/a_i^\mathrm{SM}$. Furthermore, in the case of Model $I_2$, the three intersection points (of any pair of couplings out of the three), which are controlled by the combinations $(3 a_1-5 a_2)$, $(3 a_1-5 a_3)$, and  $(a_2- a_3)$, respectively, 
occur at the same scales as in the SM running 
because the extra parts in these combinations, coming from $\Delta a_i$, cancel in the case of Model $I_2$.   
Although the intersection scales are the same between Model $I_2$ and the SM, the overall evolution of the couplings is still slightly different since each $\Delta a_i$ is nonzero in Model $I_2$. The modification manifests itself in the values of $\alpha_i^{-1}$ at each intersection point. This is obviously an accident in the 1-loop running. Going to the 2-loop level would most likely cause non-zero contributions in the relevant combinations $a_i$,
but these can be expected to be extremely suppressed and hence still would not provide an overall coupling unification.


%

So far, we have assumed that the leptoquark $H_{3L}\left(3,1,-\frac{1}{3}\right)_{321}$ is at the TeV-scale and that its companion  $H_{\overline{3}L}\left(\overline{3},1,\frac{1}{3}\right)_{321}$, the other color-triplet contained in $H_{L}(6,1,1)_{422}$, is heavy at the unification scale $M_U$, where the Pati-Salam structure emerges from the underlying NCG, and where the Pati-Salam symmetry, $G_{422D}$, breaks into the symmetry of the SM, as well. We have considered two versions of this scenario depending on whether the second Higgs doublet $\phi_2'\!\left(1,2,-\frac{1}{2}\right)_{321}$, contained in $\phi(1,2,2)_{422}$ together with the SM Higgs doublet $\phi_2\!\left(1,2,\frac{1}{2}\right)_{321}$, is light at the TeV-scale or heavy at the unification scale. We have shown that the running of the couplings is not improved in either of these cases.

Now, we will try a relatively general strategy to look for any improvement in favor of coupling unification. Instead of assuming that the second leptoquark, $H_{\overline{3}L}\left(\overline{3},1,\frac{1}{3}\right)_{321}$, and the second Higgs doublet $\phi_2'\!\left(1,2,-\frac{1}{2}\right)_{321}$ are either at the TeV-scale or the unification scale, we will let their masses float in between these scales in order to search for intervals of mass values leading to direct coupling unification. 
Note that we \textit{still} assume at this stage that there is no intermediate symmetry breaking. Therefore, the mass values of these particles are not dictated by a symmetry breaking mechanism but possibly by an another, yet to be determined, mechanism or they are picked randomly by the model by sheer accident. 
Note that we still keep the mass of our main leptoquark $H_{3L}\left(3,1,-\frac{1}{3}\right)_{321}$ at the TeV scale, required by the possible explanation for the $B$-decay anomalies. 
Then, Eq.~(\ref{equation0}) changes in the following way:\footnote{%
The threshold corrections can be safely ignored throughout this paper. The ignored terms are in the form of $-\Delta f_c(a_ i)\ln\left(1-\epsilon_c\right)$, where $\epsilon_c\equiv \dfrac{M_c-M_{\delta_c}}{M_c}$ and $c=X, Y, U$, labelling each term according to the larger mass in each logarithm in Eq.~(\ref{equation1}). $\Delta f_c(a_ i)$ are the difference between the values of corresponding combinations of $a_i$ in front of each logarithmic factor, above and below the threshold scale $M_{\delta_c}$, respectively. For instance, the correction term for the first term in the first equation in Eq.~(\ref{equation1}) becomes  $\left[ (3 a_1-5 a_2)^\mathrm{II}-(3 a_1-5 a_2)^\mathrm{I}\right]\ln\dfrac{M_X}{M_{\delta_X}}$ . From the effective field theory point of view, these terms are expected to become relevant as $M_{\delta_c}\rightarrow M_c$, or $\epsilon_c\rightarrow 0$, but then they are suppressed by a factor proportional $O(\epsilon_c)$ due to the natural logarithm, unless $\Delta f_c(a_ i)$ is enormous which is rarely the case in general, and definitely not the case for the scenarios considered in this paper.}
\begin{eqnarray}
\label{equation1}
\lefteqn{
2\pi\left[\frac{3-8 \sin^2\theta_w (M_Z)}{\alpha(M_Z)}\right]} \vphantom{\Bigg|}\cr
& \quad = &
(3 a_1-5 a_2)^\mathrm{I}\ln\frac{M_X}{M_Z}+(3 a_1-5 a_2)^{\mathrm{II}}\ln\frac{M_{Y}}{M_X}+(3 a_1-5 a_2)^{\mathrm{III}}\ln\frac{M_U}{M_{Y}} \;,
\vphantom{\Bigg|}\cr
\lefteqn{2\pi\left[\frac{3}{\alpha(M_Z)}-\frac{8}{\alpha_s (M_Z)}\right]} \vphantom{\Bigg|}\cr
& \quad = & 
(3 a_1+3 a_2-8 a_3)^\mathrm{I}\ln\frac{M_X}{M_Z}+(3 a_1+3 a_2-8 a_3)^{\mathrm{II}}\ln\frac{M_{Y}}{M_X}
+(3 a_1+3 a_2-8 a_3)^{\mathrm{III}}\ln\frac{M_U}{M_{Y}}\;,
\vphantom{\Bigg|}\cr & &
\end{eqnarray}
where $M_X$ and $M_Y$ denote, respectively, the lighter and heavier mass among the second Higgs doublet and the second leptoquark. 
Let us first consider the case where the second Higgs doublet is heavier, which we call model $T_1$ (and $T_2$ for the opposite case). 
Then, we have $M_Y\equiv M_{\phi_2'}$ and $M_X\equiv M_{H_{\overline{3}L}}$.~Therefore, in the fist interval (I) the active degrees of freedom, in addition to the SM particles, come only from our TeV-scale leptoquark $H_{3L}\left(3,1,-\frac{1}{3}\right)_{321}$. %
In the second interval (II) we add in the second leptoquark, $H_{\overline{3}L}\left(\overline{3},1,\frac{1}{3}\right)_{321}$, and finally in the third interval (III), all of the degrees of freedom are active. 
%
%
%
\begin{table}[t]
\begin{center}
{\begin{tabular}{c|l|l}
\hline
$\vphantom{\Big|}$Interval & active scalar dofs for & RG coefficients
\\
$\vphantom{\big|}$ & model $T_1$ $(T_2)$ & \\
\hline
$\vphantom{\Bigg|}$ III $(M_U-M_Y)$
& $\phi_2^{\phantom{\prime}},\; H_{3L}^{\phantom{\dagger}}$, $H_{\overline{3}L}$,\; $\phi_2'$
& $\bigl[a_{1},a_{2},a_{3}\bigr]^\mathrm{III}
  \;=\;\left[\dfrac{65}{9},-3,-\dfrac{20}{3}\right]$
\\
\hline
$\vphantom{\Bigg|}$ II $(M_Y-M_X)$
& $\phi_2^{\phantom{\prime}},\; H_{3L}^{\phantom{\dagger}},\;H_{\overline{3}L}$  
& $\bigl[a_{1},a_{2},a_{3}\bigr]^\mathrm{II}$
\\
$\vphantom{\Bigg|}$
& $\biggl(\phi_2^{\phantom{\prime}},\; H_{3L}^{\phantom{\dagger}},\; \phi_2' \biggr)$ 
& $\;=\;\left[
\dfrac{127}{18}\left(\dfrac{64}{9}\right),-\dfrac{19}{6}\left(-3\right),-\dfrac{20}{3}\left(-\dfrac{41}{6}\right)
\right]$
\\
\hline
$\vphantom{\Biggl|}$   I $(M_X-M_Z)$
& $\phi_2^{\phantom{\prime}}\;, H_{3L}^{\phantom{\dagger}}$
& $\bigl[a_{1},a_{2},a_{3}\bigr]^\mathrm{I}
  \;=\;\left[
  \dfrac{125}{18},-\dfrac{19}{6},-\dfrac{41}{6}
  \right]$
\\
\hline
\end{tabular}}
\caption{\label{xx}The Higgs content and the RG coefficients in the energy intervals for model $T_{1,2}$. Recall that all the fields above are complex.}
\end{center}
\end{table}

Using the RG coefficients given in Table~\ref{xx}, Eq.~\eqref{equation1} becomes
\begin{eqnarray}
\label{eqs1}
3270 & = & 110\,u+y-x\;,\cr
2283 &= & 66\,u-y+x\;,
\end{eqnarray}
where
\begin{equation}
u \;\equiv\; \ln\dfrac{M_U}{\mbox{GeV}}\;,\quad 
y \;\equiv\; \ln\dfrac{M_Y}{\mbox{GeV}}\;,\quad 
x \;\equiv\; \ln\dfrac{M_X}{\mbox{GeV}}\;.
\end{equation}
Solving these equations yields
\begin{equation}
u\;=\;31.55\;, \qquad y\;=\;x-200.63\;,
\end{equation}
which clearly violates our necessary condition that $u\geqslant y \geqslant x$ and hence, does not constitute a solution for our system. Similarly for model $T_2$, at which the second leptoquark is heavier than the second Higgs doublet, \textit{i.e.} $M_Y\equiv M_{H_{\overline{3}L}}$ and $M_X\equiv {\phi_2'}$, Eq.~\eqref{equation1} reads 
\begin{eqnarray}
\label{eqs2}
3270 & = & 110\,u-y+x\cr
2283 & = & 66\,u+y-x\;,
\end{eqnarray}
which yields
\begin{equation}
u\;=\;31.55\;,\qquad y\;=\;x+200.63\;,
\end{equation}
violating again the condition $u\geqslant y \geqslant x$, and thus, no acceptable solution exists in this case as well. 

Therefore, we conclude that unification of the couplings cannot be directly realized in these scenarios. There should exist at least one intermediate symmetry breaking phase between the unification scale $M_U$ and the weak scale, as we illustrate in the next section. However, we would like to note that we adopt here quite a minimalistic approach. If one becomes willing to include more degrees of freedom in the light(er) spectrum, it is quite likely that unification of the gauge couplings is realized with no intermediate symmetry breaking~\cite{Murayama:1991ah,Cox:2016epl}.

\subsection{Unification with a single intermediate scale}

\noindent
Next, we demonstrate that the unification of the couplings in Model C with a light leptoquark $H_{3L}\left(3,1,-\frac{1}{3}\right)_{321}\equiv S_1^*$ can be realized with the introduction of a single intermediate symmetry breaking step. According to this scenario, the Pati-Salam symmetric phase is intact from the unification scale $M_U$, where the Pati-Salam structure emerges from a non-commutative geometry, down to the intermediate energy scale $M_C$ ($M_Z<M_C<M_U)$, at which the $G_{422D}$ symmetry spontaneously breaks down to $G_{321}$ of the SM.

In Model C, the main field surviving down to low energies in addition to the SM fields is the leptoquark  $H_{3L}\left(3,1,-\frac{1}{3}\right)_{321}$,  originated from the complex field $H_L(6,1,1)_{422}$. The other leptoquark, $H_{\overline{3}L}\left(\overline{3},1,\frac{1}{3}\right)_{321}$, also contained in $H_L(6,1,1)_{422}$, is assumed to be heavy at the scale $M_C$. 
We consider two versions of this model, we call model $C_1$ and model $C_2$, depending on the difference based on whether the second scalar doublet, contained in the complex $\phi(1,2,2)_{422}$, remains heavy at the scale $M_C$, or it becomes light and survives down to the low energies with the other scalar doublet of the same Pati-Salam multiplet, the SM Higgs. In the latter case, the low energy section of the model can be parameterized as the 2HDM, augmented by the leptoquark $H_{3L}\left(3,1,-\frac{1}{3}\right)_{321}$.\footnote{A study of SO(10) realization of such a model and its LHC phenomenology is currently under preparation by Aydemir et.al.}

Therefore, the scalar content of the models at the TeV-scale are the same as in Eq.~\eqref{model0}, while, here, we have a different sequence of symmetry breaking. The symmetry breaking chain we consider here has been discussed in detail in our previous papers \cite{Aydemir:2015nfa, Aydemir:2016xtj}:

\begin{equation}
\label{chain}
\begin{array}{lll}
\mathrm{NCG} &\;\;\xRightarrow{\;\;\;M_U\;\;\;}\;\; G_{422D} &\;\;\underset{\langle\Delta_R\rangle}{\xrightarrow{\quad M_C\quad}}\;\; G_{321}\;,
\end{array}
\end{equation}
where the double arrow points to the symmetry emerging from the underlying NCG, while the single arrow denotes the spontaneous symmetry breaking in the usual way.

We label the energy intervals in between symmetry breaking scales
$[M_Z,M_C]$ and $[M_C,M_U]$ with Roman numerals as 
\begin{eqnarray}
\mathrm{I}  & \;:\; & [M_Z,\;M_C]\;,\quad G_{321}  \;(\mathrm{SM}) \;,\cr
\mathrm{II} & \;:\; & [M_C,\;M_U]\;,\quad  G_{422D} \;.
\label{IntervalNumber}
\end{eqnarray}
%

%

%
The generic boundary/matching conditions to be imposed on the couplings at the symmetry breaking scales are given as:
\begin{eqnarray}
M_U & \;:\; & g_L(M_U) \;=\; g_R(M_U) \;=\; g_4(M_U) \;, \vphantom{\bigg|} 
\cr
M_C & \;:\; & g_3(M_C)=g_4(M_C) \;,\quad g_2(M_C)\;=\;g_L(M_C)\;,\cr
 & \;\; & \frac{1}{g_1^2(M_C)} \;=\; \frac{1}{g_R^2(M_C)}+ \frac{2}{3}\frac{1}{g_4^2(M_C)}\;,\quad\;  g_L(M_C)\;=\;g_R(M_C),
\vphantom{\Bigg|}
\cr
M_Z & \;:\; & \frac{1}{e^2(M_Z)} \;=\; \frac{1}{g_1^2(M_Z)}+\frac{1}{g_2^2(M_Z)}\;.\vphantom{\Bigg|} 
\label{Matching}
\end{eqnarray}

\begin{table}[h]
\begin{center}
{\begin{tabular}{c|l|l}
\hline
$\vphantom{\Big|}$Interval & Higgs content for model $C_1$ $(C_2)$ & RG coefficients
\\
\hline
$\vphantom{\bigg|}$ II
& $\phi(1,2,2)_{422},\;\widetilde{\Sigma}(15,2,2)_{422},\;$ 

\\
& $\Delta_R(10,1,3)_{422},\;H_R(6,1,1)_{422}$ &$\bigl[a_{L},a_{R},a_{4}\bigr]^\mathrm{II}\;=\left[
 \dfrac{26}{3},\dfrac{26}{3},\dfrac{4}{3}
 \right]$\\
& $\Delta_L(10,3,1)_{422},\;H_L(6,1,1)_{422}$ &
\vphantom{\bigg|}
\\
\hline
$\vphantom{\Biggl|}$   I
& $\phi_2\left(1,2,\frac{1}{2}\right)_{321} 
\biggl(+\;\phi_2'\left(1,2,-\frac{1}{2}\right)_{321}\mbox{ for } C_{2} \biggr),$
& $\bigl[a_{1},a_{2},a_{3}\bigr]^\mathrm{I}$
\\
& 
$H_{3L}\!\left(3,1,-\frac{1}{3}\right)_{321}$
& $\;=\left[
 \dfrac{125}{18}\left(\dfrac{64}{9}\right),-\dfrac{19}{6}(-3),-\dfrac{41}{6}
 \right]$
$\vphantom{\Bigg|}$ 
\\
\hline
\end{tabular}}
\caption{\label{a1C}The Higgs content and the RG coefficients in the energy intervals for model $C_{1,2}$. Recall that all the fields above are complex.}
\end{center}
\end{table}

\begin{table}[b]
\begin{center}
{\begin{tabular}{c||c|c}
\hline
$\vphantom{\Big|}$
\hspace{1cm}Model \hspace{1cm}
& \hspace{1cm} $C_1$ \hspace{1cm} 
& \hspace{1cm} $C_2$ \hspace{1cm} 
\\
\hline\hline
$\vphantom{\bigg|}$ UV Symmetry
 & $G_{422D}$ & $G_{422D}$ \\
\hline
$\vphantom{\bigg|}$ $\log_{10}(M_U/\mathrm{GeV})$  &$15.7$& $15.4$ \\
$\vphantom{\bigg|}$ $\log_{10}(M_C/\mathrm{GeV})$ & $13.7$& $13.7$ \\ 
\hline
$\vphantom{\bigg|}$ $\alpha_U^{-1}$  & $36.9$& $37.0$ \\
\hline
\end{tabular}}
\caption{The predictions of models $C_1$ and $C_2$.}
\label{Results}
\end{center}
\end{table}

Using Eq.~\eqref{1looprunning} with the boundary conditions given in Eqs.~\eqref{SMboundary} and \eqref{Matching}, one obtains the relevant equations as the following.
\begin{eqnarray}
2\pi\left[\dfrac{3-8\sin^2\theta_W(M_Z)}{\alpha(M_Z)}\right]
& = & 
 \left(3a_1 -5a_2\right)^\mathrm{I}\ln\dfrac{M_C}{M_Z}
+\left(-5a_L+3a_R+2a_4\right)^\mathrm{II}\ln\dfrac{M_U}{M_C}
\;,
\vphantom{\Bigg|}
\cr
2\pi\left[\dfrac{3}{\alpha(M_Z)} - \dfrac{8}{\alpha_s(M_Z)}\right]
& = &
 \left(3a_1 + 3a_2 - 8a_3\right)^\mathrm{I}\ln\dfrac{M_C}{M_Z}
+\left(3a_L+3a_R-6a_4\right)^\mathrm{II}\ln\dfrac{M_U}{M_C}
\;,
\vphantom{\Bigg|}
\cr
& &
\end{eqnarray}
where the low energy input is collected on the left hand side. The corresponding beta coefficients for each interval are given in Table \ref{a1C}. Additionally, the unified coupling $\alpha_U$ at scale $M_U$ can be obtained from
%
\begin{eqnarray}
\label{A6x}
\dfrac{2\pi}{\alpha_U}
\;=\;  \dfrac{2\pi}{\alpha_s(M_Z)}
-\left( a_4^\mathrm{II}\;\ln\dfrac{M_U}{M_C}
+ a_3^\mathrm{I}\;\ln\dfrac{M_C}{M_Z}
\right)
\;.
\end{eqnarray}

\begin{figure}[t]
\subfigure[ ]{\includegraphics[width=7cm]{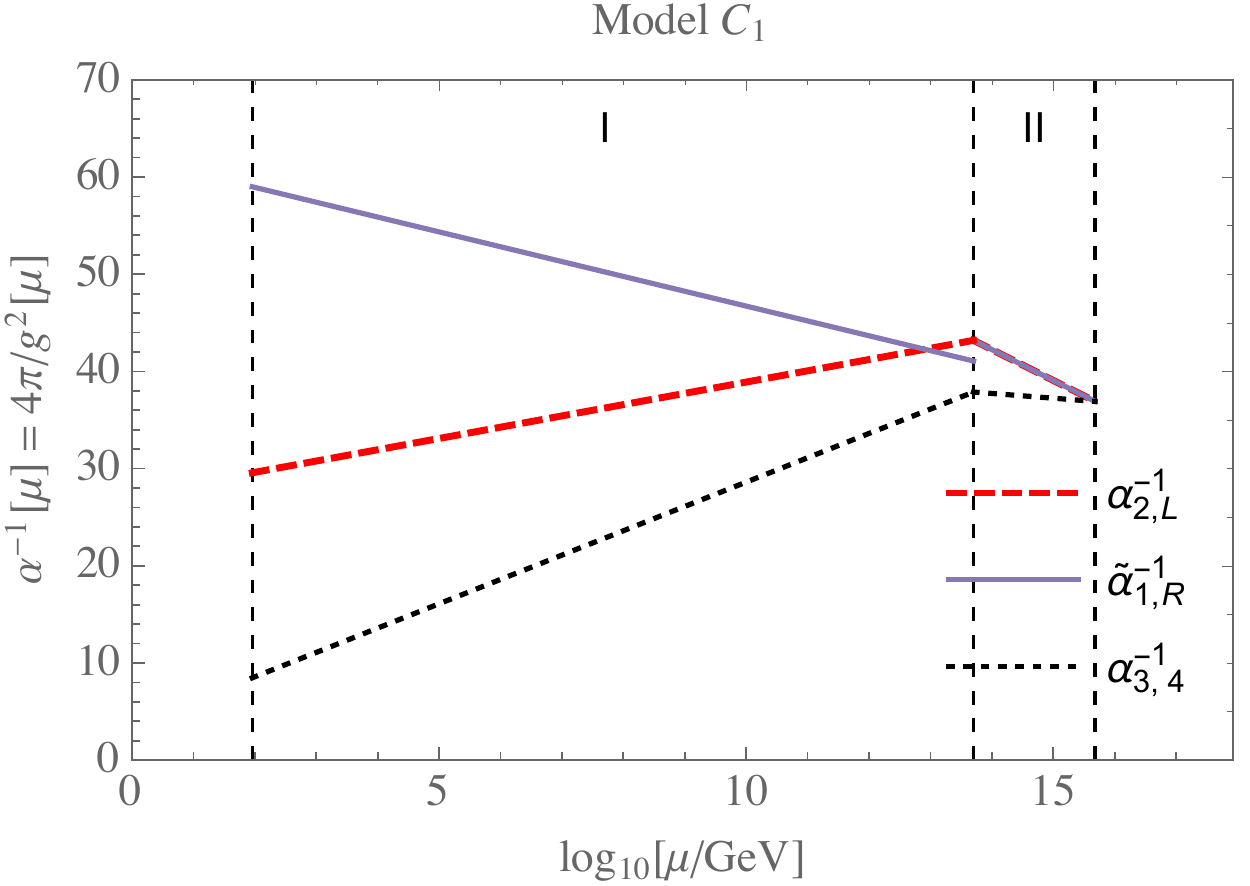}}\hspace{0.4cm}
\subfigure[ ]{\includegraphics[width=7cm]{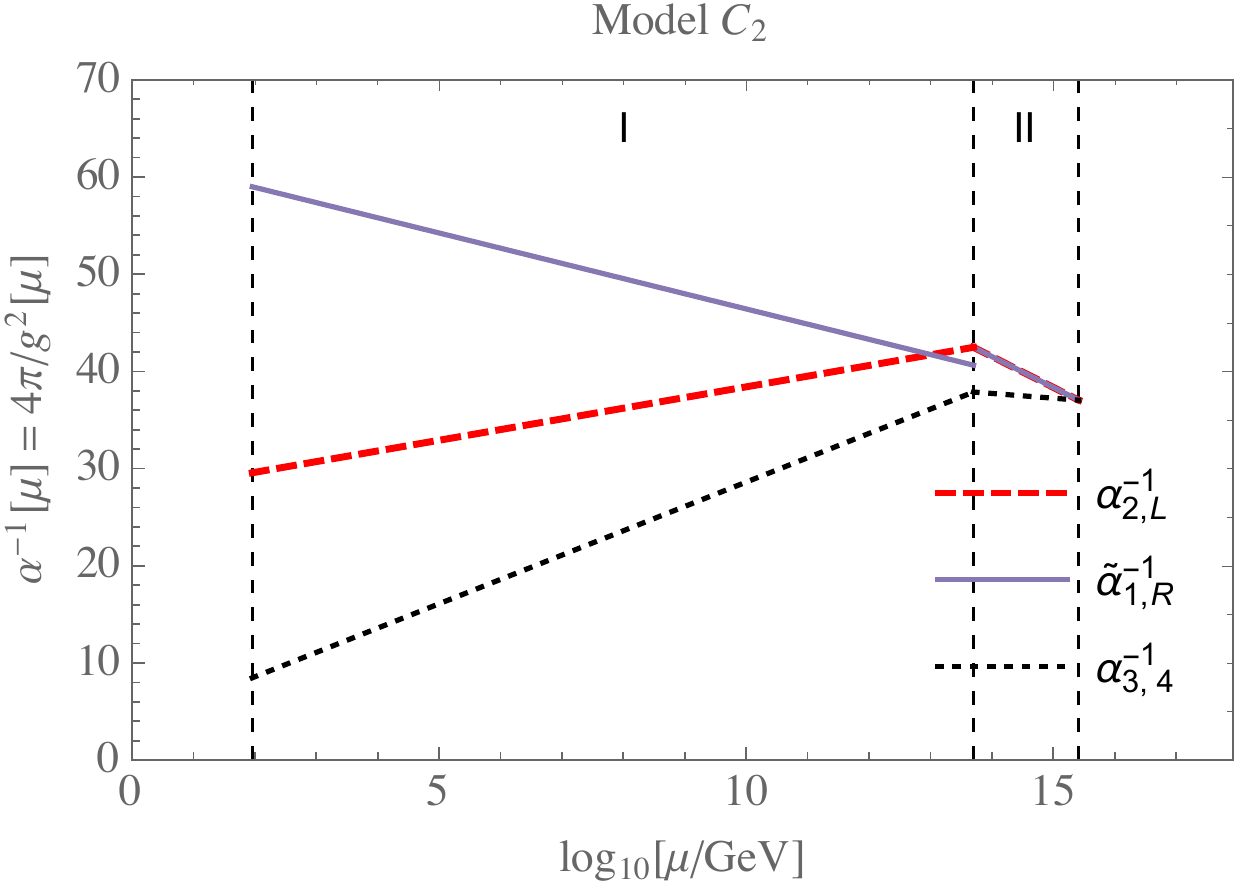}}\hspace{0.4cm}
\caption{Running of the gauge couplings for models $C_1$ and $C_2$. The vertical dotted lines correspond to the electroweak scale $M_Z$, $M_C$, and the unification scale $M_U$. Note again that $\widetilde{\alpha}^{-1}_1\equiv \frac{3}{5}\alpha^{-1}_1$.
}
\label{RGrunning2}
\end{figure}

The results are given in Table~\ref{Results} and in Fig.~\ref{RGrunning2}, where in the latter the unification of the couplings is displayed. The difference between two models is minor, as expected. In both models, the unification scales are far away from the Planck scale so that we can safely ignore gravitational effects. Moreover, the value of intermediate symmetry breaking scale, $M_C\simeq 10^{13.7}$ GeV, which is almost the same in both models, is consistent with the current bounds coming from the proton decay searches. Since this is the scale at which the Pati-Salam symmetry beaks into the SM, it determines the expected mass values for the proton-decay-mediating leptoquarks. From a naive analysis~\cite{Altarelli:2013aqa}, it can be shown that the current bounds on the proton lifetime~\cite{Takhistov:2016eqm,Patrignani:2016xqp} requires $M_C\gtrsim 10^{11}$~GeV. 

We discuss the proton stability more in the next, concluding, section, in which we address also some conceptual issues 
related to the NCG approach to the SM and its natural Pati-Salam-like completion. 
 
\section{Discussion and Conclusions}

\subsection{Discussion}


Motivated by scalar-leptoquark explanations of the recently reported $B$-decay anomalies, in this paper
we have investigated whether the required leptoquarks can be accommodated within unified
models based on NCG.
As already pointed out, such NCG-based models have the gauge structure of Pati-Salam models,
$SU(4)\times SU(2)_L\times SU(2)_R$, 
with gauge coupling unification at a single scale.
In one of the models
we did find a unique scalar leptoquark $H_{3L}(3,1,-\frac{1}{3})_{321}$, originating from the multiplet $H_L(6,1,1)_{422}$, which can potentially explain the $B$-decay anomalies, following the phenomenological analysis of \cite{Bauer:2015knc}, provided its mass is on the order of a few TeV.
Also, the unification of couplings can be realized with the inclusion of a single step of intermediate symmetry breaking. Note that $H_L(6,1,1)_{422}$ is complex and decomposes into $S_1^*\left(3,1,-\frac{1}{3}\right)_{321}$ and another field which
couples only to diquarks.  In order to avoid proton decay,  we suppress the mixing between these two scalars.  Furthermore, we assume that the field that couples only to diquarks is heavy to avoid any other phenomenological consequences. These assumptions might raise the question of naturalness, which is not different from the corresponding situation found in generic GUT models. As explained in the discussion that follows, it is an interesting feature of the NCG models that one obtains relatively light leptoquarks without obvious problems associated with proton decay.

Given the restrictive nature of the unified Pati-Salam models in the NCG framework, we find the possible scalar leptoquark $S_1^*\left(3,1,-\frac{1}{3}\right)_{321}$ explanation of the $B$-decay anomalies particularly interesting.
In particular, as pointed out in \cite{Bauer:2015knc}, this leptoquark can be used to explain not only the violation of lepton universality in $B$-decays and the enhanced rates in certain $B$-decays, but also the anomalous magnetic moment of the muon.


Note that a state with quantum numbers $\left(3,1,-\frac{1}{3}\right)_{321}$ exists in the 5 dimensional representation of $SU(5)$ as well, and therefore it could provide the required phenomenological features. However, one light 
$\left(3,1,-\frac{1}{3}\right)_{321}$
state found in a non-supersymmetric framework, does not imply unification of couplings as the $SU(5)$ GUT theory would require. The $SO(10)$ picture, on the other hand, can accommodate this in a reasonable scenario (as in Refs.~\cite{Aydemir:2015oob,Aydemir:2016qqj}) by including intermediate symmetry breaking scales.

Similarly, unification of couplings in the NCG models could be realized by imposing (at least) one intermediate scale. According to the scenario we considered in this paper, the Pati-Salam symmetry, $G_{422} = SU(4)_C \otimes SU(2)_L\otimes SU(2)_R$, emerges from an underlying non-commutative geometry at a "unification" scale $M_U$, and stays intact down to the scale $M_C$, where it is spontaneously broken to the symmetry of the SM. 
As discussed in the section 4 of this paper, there are three different Pati-Salam models in the NCG framework~\cite{Chamseddine:2013rta,Chamseddine:2015ata} depending on their scalar contents. 
However, only the $H_L(6,1,1)_{422}$ in Model C has the right coupling to left-handed fermions and it leads
to the phenomenologically preferred leptoquark. And, as we discuss in the following subsection, the usual problems
associated with proton decay can be avoided.

\subsection{Light colored scalars and proton decay}


Proton decay is not an issue of concern in our scenario, although we have a light scalar leptoquark  $H_{3L}\left(3,1,-\frac{1}{3}\right)_{321}$ (generally labeled as $S_1^*$ in the literature) in Model C, which possesses the right quantum numbers for it to couple to potentially dangerous diquark operators. The proton stability is ensured because the underlying non-commutative geometry does not allow these couplings to appear. 

Oftentimes, the possible existence of light leptoquarks is dismissed, perhaps a bit too quickly, since they can potentially lead to proton decay. Although the operators contributing to proton decay should be carefully treated or possibly turned off 
 in order not to contradict the experimental evidence, the underlying mechanism responsible for possible elimination of these operators may not be obvious at first glance. We see an example of this in Model C. In this model, the leptoquarks $H_{3R/L}\left(3,1,-\frac{1}{3}\right)_{321}$ can safely be light since their diquark couplings are turned off by the underlying noncommutative geometry and hence do not contribute to proton decay; their companions, on the other hand, $H_{\overline{3}R/L}\left(\overline{3},1,\frac{1}{3}\right)_{321}$ (\`a la the parent-multiplets $H_{R/L}(6,1,1)_{422}$), possess these couplings, and therefore the mixing between them should be small. However, by looking only at the low energy behaviour of the model, from bottom-up perspective, 
one observes only the absence of the corresponding diquark couplings. From this viewpoint, there is no indication regarding the nature of the underlying mechanism, which is the non-commutative framework in this particular case.

In general, the ordinary, non-unified, Pati-Salam-type models, have the reputation of being safe in terms of proton stability. Although this is indeed the case in terms of the gauge-boson-mediated contributions, the situation is not so trivial for the colored-scalar-mediated contributions; not every scalar sector choice guarantees the apparent stability of the proton. The most commonly chosen scalar sector in the ordinary Pati-Salam framework accommodates a global symmetry that prevents proton decay. This scalar sector consists of $\Delta_{R}\left(10,1,3\right)_{422}$ and $\Delta_{L}\left(10,3,1\right)_{422}$, in addition to the bidoublet $\phi\left(1,2,2\right)_{422}$. This is, at first glance, very similar to the scalar sector of Model C. However, this mechanism is due to the totally-symmetric nature of $\Delta_{L,R}$ under $SU(4)$, and existence of any field(s) transforming anti-symmetrically would potentially spoil this symmetry~\cite{Mohapatra:1980qe}. In fact, this is exactly what we have in Model C;  the multiplets $H_{R/L}(6,1,1)_{422}$ reside in the totally-anti-symmetric representations of $SU(4)$. Fortunately enough, the restrictive nature of the non-commutative-geometry framework happens to operate in our favor and does not allow the proton-decay-inducing diquark couplings for our leptoquark $H_{3L}\left(3,1,-\frac{1}{3}\right)_{321}$, thus offering a plausible explanation for the $B$-decay anomalies.     
 

\subsection{NCG's Pros and Cons}

\noindent

In this paper we have discussed yet another phenomenological aspect of the hidden non-commutative structure 
behind the Standard Model (and its natural Pati-Salam-like completion), that
has been uncovered by Connes, Chamseddine and collaborators \cite{Chamseddine:1996zu, Chamseddine:2012sw}.
Even though such non-commutative framework has been pointed out for some time, in light of the LHC's discovery of the Higgs boson, and the fact that the NCG of the SM is precisely relevant for the SM with the Higgs boson, we were motivated to explore other phenomenological implications of such a non-commutative approach \cite{Aydemir:2013zua,Aydemir:2014ama,Aydemir:2015nfa,Aydemir:2016xtj}. 

Note that the NCG of the SM might appear peculiar from the canonical effective field theory point of view, because of the way the SM action is rewritten, in a Fujikawa-like form familiar from the study of chiral anomalies in gauge theories, which introduces a scale, not a Wilsonian cut-off, and which in turn is fixed by the existence of one overall coupling, thus implying the unification of the three SM couplings at a GUT scale without GUT degrees of freedom.
Furthermore, the NCG scheme hides a left-right symmetric structure which can be naturally broken to the canonical SM. 
Note that this is also an unusual left-right completion.  First, this non-commutative completion has a GUT-like unification (like the NCG of the SM), which simply does not exist for the canonical left-right symmetric models. Second, this non-commutative left-right completion of the SM is quite constrained compared to the already existing effective field theory literature on the subject.
As emphasized in the present paper, such unified Pati-Salam structure can lead to a unique scalar leptoquark which may potentially explain the $B$-decay anomalies, provided its mass is around a few TeV, without causing any problems with proton decay.

One might wonder whether the NCG of the SM and the unified Pati-Salam models can have a reasonable UV completion.
In particular, where could the NCG originate from a UV point of view? Here, in conclusion, we offer a few comments on this
important question.
We note that if one views quantum gravity coupled to Standard Model-like matter as having origins in string theory, then one can convincingly argue for an intrinsic non-commutative structure  in that context
\cite{Freidel:2017nhg, Freidel:2017wst, Freidel:2017xsi, Freidel:2016pls, Freidel:2015pka, Freidel:2014qna, Freidel:2013zga}.
This work also suggests that the low energy theory (such as the one described by the Standard Model and its Pati-Salam-like completion) is intrinsically non-commutative. In order to obtain the NCG of the SM and its Pati-Salam completion, from such a UV point of view, what is needed is just some suitable fermionic degrees of freedom, propagating in a background of induced gravitational and gauge degrees of freedom (of the SM or Pati-Salam type). Then after integrating over the fermionic zero modes the induced action of the Fujikawa form, needed by the NCG structure of the SM, naturally emerges.
Note that the usual discussion of hierarchy of scales as well as the issue of decoupling of scales is different in this framework, which in principle allows for mixing between short distance and long distance degrees of freedom, thus shedding new light on the questions of naturalness and the hierarchy problem.




\section*{Acknowledgments}

We thank Chris Hill, Paul Oehlmann, Nobuchika Okada, Leo Piilonen and
Mark Pitt for discussions and Walter D. van Suijlekom for useful
communications.~UA is supported by the National Natural Science Foundation of China (NSFC) under Grant No.~11505067.  DM
thanks the Julian Schwinger Foundation for support.~CS is supported in
part by the International Postdoctoral Fellowship funded by China
Postdoctoral Science Foundation, and is grateful for the support from
JiJi Fan, Marcelo Gleiser, and Devin Walker.


\appendix
\section{Leptoquarks}

In this appendix we review the derivation of the
most generic Lagrangian for scalar and vector leptoquark
interactions with Standard Model (SM) fermions.
We follow the notation of ref.~\cite{Dorsner:2016wpm},
which updates ref.~\cite{Buchmuller:1986zs} with the inclusion of leptoquarks
and interactions that involve the right-handed neutrino.~All quantum number assignments refer to those in $G_{321}=SU(3)_C\times SU(2)_L\times U(1)_Y$ of the SM.


At low-energies,
we assume the fermionic content of the SM plus right-handed neutrinos:
\begin{equation}
\begin{array}{lll}
Q_L^i \,=\, \left(3,2,+\dfrac{1}{6}\right)\,,\qquad &
d_R^i \,=\, \left(3,1,-\dfrac{1}{3}\right)\,,\qquad &
u_R^i \,=\, \left(3,1,+\dfrac{2}{3}\right)\,, \\
L_L^i \,=\, \left(1,2,-\dfrac{1}{2}\right)\,,\qquad &
e_R^i \,=\, \left(1,1,-1\right)\,,\qquad &
\nu_R^i \,=\, \left(1,1,0\right)\,.
\end{array}
\end{equation}
where $i=1,2,3$ is the generation index.
From this set of fermions, let us construct the most general quark-lepton bilinears.
First, the scalars with fermion number $F=3B+L=0$ are
\begin{equation}
\begin{array}{rlrl}
\overline{u_R^{i}}L_L^j & =\, \left(\bar{3},2,-\dfrac{7}{6}\right)\,,\qquad &
\overline{d_R^{i}}L_L^j & =\, \left(\bar{3},2,-\dfrac{1}{6}\right)\,,\\
\overline{Q_L^i}e_R^{j} & =\, \left(\bar{3},2,-\dfrac{7}{6}\right)\,,\qquad &
\overline{Q_L^i}\nu_R^{j} & =\, \left(\bar{3},2,-\dfrac{1}{6}\right)\,,\\
\end{array}
\end{equation}
and their hermitian conjugates.
Note that the bilinears in the same column above share the same 
$SU(3)_C\times SU(2)_L\times U(1)_Y$ 
quantum numbers.
The scalars with fermion number $F=2$ are
\begin{equation}
\begin{array}{rlrlrl}
\overline{Q_L^{Ci}}\epsilon L_L^j & =\, \left(3,1,-\dfrac{1}{3}\right)\,,\qquad &
\overline{u_R^{Ci}}e_R^j & =\, \left(3,1,-\dfrac{1}{3}\right)\,,\qquad &
\overline{d_R^{Ci}}\nu_R^j & =\, \left(3,1,-\dfrac{1}{3}\right)\,,\\
\overline{Q_L^{Ci}}\epsilon \vec{\tau} L_L^j & =\, \left(3,3,-\dfrac{1}{3}\right)\,,\qquad &
\overline{d_R^{Ci}}e_R^j & =\, \left(3,1,-\dfrac{4}{3}\right)\,,\qquad &
\overline{u_R^{Ci}}\nu_R^j & =\, \left(3,1,+\dfrac{2}{3}\right)\,,\\
\end{array}
\end{equation}
and their hermitian conjugates, where $\psi^C = C\overline{\psi}^\mathrm{T}$,
and $\epsilon = i\tau_2$ in $SU(2)_L$ isospin space which contracts the isospin indices of two isospinors.
$\vec{\tau} = (\tau_1,\tau_2,\tau_3)$ are the Pauli matrices.
Note that the bilinears in the first row above share the same quantum numbers.
The $F=0$ vectors are
\begin{equation}
\begin{array}{rlrlrl}
\overline{Q_L^{i}}\gamma^\mu L_L^j & =\, \left(\bar{3},1,-\dfrac{2}{3}\right)\,,\qquad &
\overline{d_R^{i}}\gamma^\mu e_R^j & =\, \left(\bar{3},1,-\dfrac{2}{3}\right)\,,\qquad &
\overline{u_R^{i}}\gamma^\mu \nu_R^j & =\, \left(\bar{3},1,-\dfrac{2}{3}\right)\,,\\
\overline{Q_L^{i}}\vec{\tau}\gamma^\mu L_L^j & =\, \left(\bar{3},3,-\dfrac{2}{3}\right)\,,\qquad &
\overline{u_R^{i}}\gamma^\mu e_R^j & =\, \left(\bar{3},1,-\dfrac{5}{3}\right)\,,\qquad &
\overline{d_R^{i}}\gamma^\mu \nu_R^j & =\, \left(\bar{3},1,+\dfrac{1}{3}\right)\,,\\
\end{array}
\end{equation}
and their hermitian conjugates.
Again, the bilinears in the first row share the same quantum numbers.
The $F=2$ vectors are
\begin{equation}
\begin{array}{rlrl}
\overline{d_R^{Ci}}\gamma^\mu L_L^j & =\, \left(3,2,-\dfrac{5}{6}\right)\;,
\qquad &
\overline{u_R^{Ci}}\gamma^\mu L_L^j & =\, \left(3,2,+\dfrac{1}{6}\right)\;,
\\
\overline{Q_L^{Ci}}\gamma^\mu e_R^j & =\, \left(3,2,-\dfrac{5}{6}\right)\;,
\qquad &
\overline{Q_L^{Ci}}\gamma^\mu \nu_R^j & =\, \left(3,2,+\dfrac{1}{6}\right)\;,\\
\end{array}
\end{equation}
and their hermitian conjugates, the bilinears in the same columns sharing the same quantum numbers.

Introducing one scalar or vector leptoquark for each possible combination of quantum numbers, the most general Lagrangian which couple leptoquarks to
these bilinears can be written down as
%
\begin{equation}
\mathcal{L} \;=\; \mathcal{L}_{F=2} + \mathcal{L}_{F=0}\;,
\end{equation}
where
\begin{eqnarray}
\mathcal{L}_{F=2} &=& 
\biggl[  
 y_{1\,ij}^{LL}\Bigl(\overline{Q_L^{Ci}}\epsilon L_L^{j} \Bigr)
+y_{1\,ij}^{RR}\Bigl(\overline{u_R^{Ci}} e_R^{j} \Bigr)
+y_{1\,ij}^{\overline{RR}}\Bigl(\overline{d_R^{Ci}} \nu_R^{j} \Bigr)
\biggr] S_1
\nonumber\\
&& 
+\;\widetilde{y}_{1\,ij}^{RR}\Bigl(\overline{d_R^{Ci}} e_R^{j} \Bigr) \widetilde{S}_1
+\overline{y}_{1\,ij}^{\overline{RR}}\Bigl(\overline{u_R^{Ci}} \nu_R^{j}\Bigr)
\overline{S}_1
+y_{3\,ij}^{LL}\Bigl(\overline{Q_L^{Ci}} \epsilon \vec{\tau} L_L^{j} \Bigr) \vec{S}_3 
\nonumber\\
&& 
+\biggl[
 x_{2\,ij}^{RL}\Bigl(\overline{d_R^{Ci}} \gamma^\mu L_L^{j} \Bigr)
+x_{2\,ij}^{LR}\Bigl(\overline{Q_L^{Ci}} \gamma^\mu e_R^{j} \Bigr)
\biggr] \epsilon V_{2\mu}
\nonumber\\
&& 
+\biggl[
 \widetilde{x}_{2\,ij}^{RL}\Bigl(\overline{u_R^{Ci}} \gamma^\mu L_L^{j} \Bigr) 
+\widetilde{x}_{2\,ij}^{\overline{LR}}\Bigl(\overline{Q_L^{Ci}} \gamma^\mu \nu_R^{j} \Bigr) 
\biggr] \epsilon \widetilde{V}_{2\mu}
+ h.c. \;,
\label{lqlag:f2}
\\ 
& & \vphantom{\big|} \cr
\mathcal{L}_{F=0}
&=&
\biggl[
 y_{2\,ij}^{RL}\Bigl(\overline{u_R^i}L_L^j \Bigr)\epsilon
+y_{2\,ij}^{LR}\Bigl(\overline{Q_L^i}e_R^j \Bigr)
\biggr] R_2
\nonumber\\
&&
+\biggl[
 \widetilde{y}_{2\,ij}^{RL}\Bigl(\overline{d_R^i} L_L^j \Bigr)\epsilon
+\widetilde{y}_{2\,ij}^{\overline{LR}}\Bigl(\overline{Q_L^i}\nu_R^j \Bigr)
\biggr] \widetilde{R}_2
\nonumber\\
&&
+\biggl[
 x_{1\,ij}^{LL}\Bigl(\overline{Q_L^i}\gamma^\mu L_L^j \Bigr) 
+x_{1\,ij}^{RR}\Bigl(\overline{d_R^i}\gamma^\mu e_R^j \Bigr) 
+x_{1\,ij}^{\overline{RR}}\Bigl(\overline{u_R^i}\gamma^\mu \nu_R^j \Bigr)
\biggr] U_{1\mu}
\nonumber\\
&&
+\;\widetilde{x}_{1\,ij}^{RR}
\Bigl( \overline{u_R^i}\gamma^\mu e_R^j \Bigr) \widetilde{U}_{1\mu}
+\overline{x}_{1\,ij}^{\overline{RR}}
\Bigl( \overline{d_R^i}\gamma^\mu \nu_R^j \Bigr) \overline{U}_{1\mu}
+x_{3\,ij}^{LL} \Bigl( \overline{Q_L^i} \vec{\tau}\gamma^\mu L_L^j
\Bigr) \vec{U}_{3\mu} 
+h.c. \;.
\cr
& &
\label{lqlag:f0}
\end{eqnarray}
%
Here, the couplings between the
vector/scalar leptoquarks and the fermions are denoted respectively by $x$ and $y$.
The first subscript indicates the isospin of the leptoquark and the latter two the generations of the fermions involved. The superscripts denote the chiralities of the fermions, with a bar placed over the chiralities when a right-handed neutrino is involved.\footnote{%
In order to completely match the definitions of ref.~\cite{Dorsner:2016wpm},
replace $x_{2\,ij}^{RL}$ by $-x_{2\,ij}^{RL}$, and $y_{2\,ij}^{LR}$ by 
$y_{2\,ij}^{LR*}$ (complex conjugate).
}

Thus, six scalar leptoquarks
$S_1$, $\widetilde{S}_1$, $\overline{S}_1$, $S_3$, $R_2$, $\widetilde{R}_2$,
and six vector leptoquarks
$U_1$, $\widetilde{U}_1$, $\overline{U}_1$, $U_3$, $V_2$, $\widetilde{V}_2$
are introduced.
Of these, $\overline{S}_1$ and $\overline{U}_1$ were absent in ref.~\cite{Buchmuller:1986zs} since the only lepton they couple to is the right-handed neutrino. 
The quantum numbers of these twelve fields are listed in Table~\ref{LQtable}.
These leptoquarks can also have $SU(3)_C\times SU(2)_L\times U(1)_Y$ invariant
couplings to di-quarks as shown in ref.~\cite{Dorsner:2016wpm} but we will not 
discuss them here.

Note that the subscripts of the leptoquark fields indicate their isospin.
So $R_2$, $\widetilde{R}_2$, $V_2$, and $\widetilde{V}_2$ have two components each, while $S_3$ and $U_3$ have three components each.
These can be written out as
\begin{equation}
\begin{array}{rlrlrl}
R_2 & =\,\begin{bmatrix} R_2^{(5/3)} \\ R_2^{(2/3)} \end{bmatrix}\,,\qquad &
\tilde{R}_2 & =\,\begin{bmatrix} \tilde{R}_2^{(2/3)} \\ \tilde{R}_2^{(-1/3)} \end{bmatrix}\,,\qquad &
S_3 & =\,\begin{bmatrix} S_3^{(4/3)} \\ S_3^{(1/3)} \\ S_3^{(-2/3)} \end{bmatrix}\,,
\\
V_2 & =\,\begin{bmatrix} V_2^{(4/3)} \\ V_2^{(1/3)} \end{bmatrix}\,,\qquad &
\tilde{V}_2 & =\,\begin{bmatrix} \tilde{V}_2^{(1/3)} \\ \tilde{V}_2^{(-2/3)} \end{bmatrix}\,,\qquad &
U_3 & =\,\begin{bmatrix} U_3^{(5/3)} \\ U_3^{(2/3)} \\ U_3^{(-1/3)} \end{bmatrix}\,,
\\
\end{array}
\end{equation}
where the superscripts in parentheses indicate the electromagnetic charge.
Expanding out eqs.~(\ref{lqlag:f2}) and (\ref{lqlag:f0}) in terms of these component fields, we find
%
\begin{eqnarray}
\mathcal{L}_{F=2} &=& 
\biggl[
 y_{1\,ij}^{LL}\Bigl(\overline{u_L^{Ci}}e_L^j - \overline{d_L^{Ci}}\nu_L^j \Bigr)
+y_{1\,ij}^{RR}\Bigl(\overline{u_R^{Ci}}e_R^j \Bigr)
+y_{1\,ij}^{\overline{RR}}\Bigl(\overline{d_R^{Ci}}\nu_R^j \Bigr)
\biggr] S_1^{(1/3)}
\cr
&& 
+\;\widetilde{y}_{1\,ij}^{RR}\Bigl(\overline{d_R^{Ci}} e_R^{j} \Bigr) \widetilde{S}_1^{(4/3)}
+\overline{y}_{1\,ij}^{\overline{RR}}\Bigl(\overline{u_R^{Ci}} \nu_R^{j}\Bigr)
\overline{S}_1^{(-2/3)}
\cr
&&
+\;y_{3\,ij}^{LL}
\biggl[
-\sqrt{2}\Bigl(\overline{d_L^{Ci}}e_L^j\Bigr) S_3^{(4/3)}
-\Bigl(\overline{u_L^{Ci}}e_L^j+\overline{d_L^{Ci}}\nu_L^j\Bigr) S_3^{(1/3)}
+\sqrt{2}\Bigl(\overline{u_L^{Ci}}\nu_L^j\Bigr) S_3^{(-2/3)} 
\biggr]
\nonumber\\
&&
-\biggl[
 x_{2\,ij}^{RL}\Bigl(\overline{d_R^{Ci}}\gamma^\mu e_L^j\Bigr)
+x_{2\,ij}^{LR}\Bigl(\overline{d_L^{Ci}}\gamma^\mu e_R^j\Bigr) 
\biggr] V_{2\mu}^{(4/3)}
+\biggl[
 x_{2\,ij}^{RL}\Bigl(\overline{d_R^{Ci}}\gamma^\mu \nu_L^j\Bigr)
+x_{2\,ij}^{LR}\Bigl(\overline{u_L^{Ci}}\gamma^\mu e_R^j\Bigr) 
\biggr] V_{2\mu}^{(1/3)}
\nonumber\\
&&
-\biggl[
 \widetilde{x}_{2\,ij}^{RL}\Bigl(\overline{u_R^{Ci}}\gamma^\mu e_L^j\Bigr)
+\widetilde{x}_{2\,ij}^{\overline{LR}}\Bigl(\overline{d_L^{Ci}}\gamma^\mu \nu_R^j\Bigr) 
\biggr] \widetilde{V}_{2\mu}^{(1/3)}
+\biggl[
 \widetilde{x}_{2\,ij}^{RL}\Bigl(\overline{u_R^{Ci}}\gamma^\mu \nu_L^j\Bigr)
+\widetilde{x}_{2\,ij}^{\overline{LR}}\Bigl(\overline{u_L^{Ci}}\gamma^\mu \nu_R^j\Bigr) 
\biggr] \widetilde{V}_{2\mu}^{(-2/3)}
\nonumber\\
&&
+\;h.c.\;,\vphantom{\bigg|}
\label{LQcomponentcouplings2}
\\
& & \vphantom{M} \cr
%
{\cal L}_{F=0}
&=&
\biggl[
-y_{2\,ij}^{RL}\Bigl(\overline{u_R^i}e_L^j\Bigr)
+y_{2\,ij}^{LR}\Bigl(\overline{u_L^i}e_R^j\Bigr)
\biggr] R_2^{(5/3)}
+\biggl[
 y_{2\,ij}^{RL}\Bigl(\overline{u_R^i}\nu_L^j\Bigr)
+y_{2\,ij}^{LR}\Bigl(\overline{d_L^i}e_R^j\Bigr)
\biggr] R_2^{(2/3)}
\nonumber\\
&&
+ 
\biggl[
-\widetilde{y}_{2\,ij}^{RL}\Bigl(\overline{d_R^i}e_L^j \Bigr)
+\widetilde{y}_{2\,ij}^{\overline{LR}}\Bigl(\overline{u_L^i}\nu_R^j \Bigr)
\biggr] \widetilde{R}_2^{(2/3)} 
+\biggl[
 \widetilde{y}_{2\,ij}^{RL}\Bigl(\overline{d_R^i}\nu_L^j \Bigr)
+\widetilde{y}_{2\,ij}^{\overline{LR}}\Bigl(\overline{d_L^i}\nu_R^j \Bigr)
\biggr] \widetilde{R}_2^{(-1/3)}
\nonumber\\
&&
+\biggl[
 x_{1\,ij}^{LL}
 \Bigl(\overline{u_L^i}\gamma^\mu \nu_L^j 
     + \overline{d_L^i}\gamma^\mu e_L^j 
 \Bigr)
+x_{1\,ij}^{RR}
 \Bigl(\overline{d_R^i}\gamma^\mu e_R^j \Bigr) 
+x_{1\,ij}^{\overline{RR}}
 \Bigl(\overline{u_R^i}\gamma^\mu \nu_R^j \Bigr) 
\biggr]U_{1\mu}^{(2/3)}
\nonumber\\
&&
+\;\widetilde{x}_{1\,ij}^{RR}
\Big(\overline{u_R^i}\gamma^\mu e_R^j\Bigr) \widetilde{U}_{1\mu}^{(5/3)}
+\;\overline{x}_{1\,ij}^{\overline{RR}}
\Big(\overline{d_R^i}\gamma^\mu \nu_R^j\Bigr) \overline{U}_{1\mu}^{(-1/3)}
\nonumber\\
&&
+x_{3\,ij}^{LL}
\biggl[
 \sqrt{2}\Bigl(\overline{u_L^i}\gamma^\mu e_L^j \Bigr) U_{3\mu}^{(5/3)}
+\Bigl(\overline{u_L^i}\gamma^\mu \nu_L^j 
               -\overline{d_L^i}\gamma^\mu e_L^j \Bigr) U_{3\mu}^{(2/3)}
+\sqrt{2}\Bigl(\overline{d_L^i}\gamma^\mu \nu_L^j \Bigr) U_{3\mu}^{(-1/3)} 
\biggr] 
\cr
&&
+\;h.c.\;.\vphantom{\bigg|}    
\label{LQcomponentcouplings0}
\end{eqnarray}
%
where we have added superscripts indicating the electromagnetic charges to the
isosinglet fields also.



\begin{table}[ht]
\begin{center}
\begin{tabular}{|c||c|c|c|c|c|}
\hline
leptoquark & spin & $F=3B+L$ & Quantum Numbers & $Q_{em}=I_3^L+Y$ & Couples to \\
\hline\hline
$S_1$ & 0 & $-2$ & 
$\left(\overline{3},1,\dfrac{1}{3}\right)$ & $\dfrac{1}{3}$ & $LL$, $RR$ \\
\hline
$\widetilde{S}_1$ & 0 & $-2$ & 
$\left(\overline{3},1,\dfrac{4}{3}\right)$ & $\dfrac{4}{3}$ & $RR$ \\
\hline
$\overline{S}_1$ & 0 & $-2$ & 
$\left(\overline{3},1,-\dfrac{2}{3}\right)$ & $-\dfrac{2}{3}$ & $\overline{RR}$ \\
\hline
$S_3$ & 0 & $-2$ & 
$\left(\overline{3},3,\dfrac{1}{3}\right)$ & $\left(\dfrac{4}{3},\dfrac{1}{3},-\dfrac{2}{3}\right)$ & $LL$ \\
\hline
$R_2$ & 0 & $0$ & 
$\left(3,2,\dfrac{7}{6}\right)$ & $\left(\dfrac{5}{3},\dfrac{2}{3}\right)$ &  $RL$, $LR$ \\
\hline
$\widetilde{R}_2$ & 0 & $0$ & 
$\left(3,2,\dfrac{1}{6}\right)$ & $\left(\dfrac{2}{3},-\dfrac{1}{3}\right)$ & $RL$, $\overline{LR}$  \\
\hline\hline
$U_1$ & 1 & $0$ & 
$\left(3,1,\dfrac{2}{3}\right)$ & $\dfrac{2}{3}$ & $LL$, $RR$  \\
\hline
$\widetilde{U}_1$ & 1 & $0$ & 
$\left(3,1,\dfrac{5}{3}\right)$ & $\dfrac{5}{3}$ & $RR$ \\
\hline
$\overline{U}_1$ & 1 & $0$ & 
$\left(3,1,-\dfrac{1}{3}\right)$ & $-\dfrac{1}{3}$ & $\overline{RR}$ \\
\hline
$U_3$ & 1 & $0$ & 
$\left(3,3,\dfrac{2}{3}\right)$ & $\left(\dfrac{5}{3},\dfrac{2}{3},-\dfrac{1}{3}\right)$ & $LL$ \\
\hline
$V_2$ & 1 & $-2$ & 
$\left(3,2,\dfrac{5}{6}\right)$ & $\left(\dfrac{4}{3},\dfrac{1}{3}\right)$ & $RL$, $LR$ \\
\hline
$\widetilde{V}_2$ & 1 & $-2$ & 
$\left(3,2,-\dfrac{1}{6}\right)$ & $\left(\dfrac{1}{3},-\dfrac{2}{3}\right)$ & $RL$, $\overline{LR}$ \\
\hline
\end{tabular}
\caption{Quantum numbers of scalar and vector leptoquarks with
$SU(3)_C\times SU(2)_L\times U(1)_Y$ invariant couplings to quark-lepton
pairs.  $F=3B+L$ is the fermion number and the hypercharge $Y$ 
is normalized so that $Q_{\rm em}=I_3^L+Y$.
The final column indicates the chiralities of the quark and lepton fields that each leptoquark couples to, with the first letter of the pairs indicating the chirality of the quark field and the second letter indicating that of the lepton field. Pairs with a bar over them indicate that the right-handed lepton involved in the pair is a right-handed neutrino.
}
\end{center}
\label{LQtable}
\end{table}


\newpage
\bibliographystyle{JHEP}
\bibliography{NCGflavor}

\end{document}